\documentclass[fleqn,usenatbib]{mnras}
\usepackage{newtxtext,newtxmath}
\usepackage[T1]{fontenc}

\usepackage{graphicx}
\usepackage{amsmath}
\usepackage{xfrac}
\usepackage{MnSymbol}

\defcitealias{Ellis:1984uka}{EB84}
\defcitealias{Secrest:2020has}{S21}
\defcitealias{Secrest:2022uvx}{S22}

\title[Clustering of CatWISE2020 quasars]{Clustering properties of the CatWISE2020 quasar catalogue and their impact on the cosmic dipole anomaly}

\author[von Hausegger et al.]{Sebastian von Hausegger,$^{1}$\thanks{E-mail: sebastian.vonhausegger@physics.ox.ac.uk (SvH)}
Nathan Secrest,$^{2}$
Harry Desmond,$^{3}$
Mohamed Rameez,$^{4}$
Roya Mohayaee~$^{5}$\newauthor
and Subir Sarkar~$^{1}$
\\ \\
$^{1}$ Rudolf Peierls Centre for Theoretical Physics,\unpenalty~University of Oxford,\unpenalty~Parks Road,\unpenalty~Oxford OX1 3PU,\unpenalty~UK\\
$^{2}$ U.S.\ Naval Observatory,\unpenalty~3450 Massachusetts Ave NW,\unpenalty~Washington,\unpenalty~DC 20392-5420,\unpenalty~USA\\
$^{3}$ Institute for Cosmology \& Gravitation,\unpenalty~University of Portsmouth,\unpenalty~Dennis Sciama Building,\unpenalty~Portsmouth PO1 3FX,\unpenalty~UK\\
$^{4}$ Department of High Energy Physics,\unpenalty~Tata Institute of Fundamental Research,\unpenalty~Homi Bhabha Road,\unpenalty~Mumbai 400005,\unpenalty~India\\
$^{5}$ Institut d'Astrophysique,\unpenalty~CNRS/Sorbonne Universit\'e,\unpenalty~98 bis Bld Arago,\unpenalty~Paris 75014,\unpenalty~France
}

\begin{document}
\label{firstpage}
\pagerange{\pageref{firstpage}--\pageref{lastpage}}
\maketitle

\begin{abstract}
The cosmic dipole anomaly --- the mismatch between the dipole anisotropy observed in the sky distribution of cosmologically distant sources and that expected due to our local motion with respect to the cosmic microwave background --- poses a serious challenge to the Cosmological Principle upon which the standard model of cosmology rests. Accurate measurement of the dipole ($\ell=1$) depends crucially on having control over other large-scale power ($\ell > 1$) so as to avoid biases, in particular that potentially caused by correlations among multipoles due to incomplete sky coverage, and  that due to local source clustering. Currently, the most significant evidence for the cosmic dipole anomaly comes from the sample of 1.6~million mid-infrared quasars derived from the CatWISE2020 catalogue. We analyse the clustering properties of this sample by inferring the large-scale multipoles in real space, and compute the angular power spectrum on small scales to test for agreement with $\Lambda$CDM. Having accounted for the known trend of the quasar number counts with ecliptic latitude, we find that all other large-scale power is in fact consistent with noise, that there is in particular no evidence for an octupole ($\ell=3$) in the data, and also that the expected clustering dipole is marginal. Our results thus reaffirm the anomalously high dipole in the distribution of quasars.
\end{abstract}


\section{Introduction}
The Cosmological Principle asserts that the Universe is (statistically) isotropic and homogeneous when averaged on large scales. This is the foundational assumption of Friedmann-Lema\^{\i}re-Robertson-Walker (FLRW) cosmologies, which include the current standard ``concordance'' Lambda cold dark matter ($\Lambda$CDM) model. The real Universe is, however, inhomogeneous, with local departures from the Hubble flow, viz. peculiar motions on small scales. This was predicted by \citet{Stewart:1967ve} to give rise to a dipole anisotropy in the Cosmic Microwave Background (CMB) and the subsequent detection of the CMB dipole at the expected amplitude of $\beta \equiv v/c \sim 10^{-3}$ proved to be consistent with this kinematic hypothesis. 
Later, when primordial fluctuations at a much smaller level of $\sim10^{-5}$ were detected in the CMB, it was noted that the same aberration  effect should induce subtle correlations between the higher multipoles \citep{Challinor:2002zh}, and this too was detected at the expected amplitude \citep{Planck:2013kqc,Ferreira:2020aqa,Saha:2021bay}, thus cementing the kinematic interpretation. 

Subsequently, \citet{Ellis:1984uka} noted that there should be a concomitant dipole anisotropy in the sky map of cosmologically distant objects in a flux-limited catalogue, and that this provides a powerful consistency test of the FLRW assumption. In this standard framework, the Solar System barycentre is moving relative to the cosmic rest frame in which the CMB dipole should vanish, at $v = 369.82 \pm 0.11$~km\,s$^{-1}$ towards Galactic coordinates $l = (264.021 \pm 0.011)^\circ, b = (48.243 \pm 0.005)^\circ$ \citep{Planck:2018nkj}. Since the distribution of matter should also be isotropic in the cosmic rest frame, our peculiar velocity inferred from the dipole anisotropy of cosmologically distant radio sources and quasars should match this value. Any disagreement would be incompatible with an FLRW-based cosmology such as $\Lambda$CDM, in which all inferences, including that of an accelerating expansion rate, rely on large-scale isotropy.

The Ellis \& Baldwin (\citetalias{Ellis:1984uka}) test became possible only at the turn of the millennium after sufficiently large samples of sources became available.\footnote{For a recent review of the history of \citetalias{Ellis:1984uka} tests, see \citet{Secrest:2025nbt}.} First, \citet{Blake:2002gx} analysed radio galaxy samples constructed from the NRAO~VLA~Sky~Survey \citep[NVSS;][]{1998AJ....115.1693C} of size  $\mathcal{O}(10^5)$ at increasing flux density cuts and reported that the dipole amplitude and direction are, on average as the flux cut is increased, consistent to within $1.5\sigma$ with the expected kinematic dipole. This consistency, however, is guaranteed for any sample arbitrarily cut until source counts sufficiently limit statistical power. As emphasized by \citet{Secrest:2025akp}, the sample employed by \citet{Blake:2002gx} with the greatest statistical power in fact shows a $2.2\sigma$ deviation from the kinematic expectation. It is unsurprising therefore that a reanalysis by \citet{Singal:2011dy} claimed that the amplitude is larger than expected, although the direction is indeed consistent with that of the CMB dipole. This was confirmed by \citet{Rubart:2013tx} and \citet{Colin:2017juj} who included data from other radio surveys like the Westerbork~Northern~Sky~Survey \citep[WENSS;][]{1997A&AS..124..259R} and the Sydney~University~Molonglo~Sky~Survey \citep[SUMSS;][]{2003MNRAS.342.1117M} to increase sky coverage. The significance of the discrepancy has been, however, limited to the $\sim3\sigma$ level, mainly due to sample size. There is also concern about possible contamination by the ``clustering dipole'' due to  nearby structure \citep[e.g.][]{Gibelyou:2012ri} -- a recurring question in the context of matter dipole measurements using radio sources, which rarely have measured redshifts \citep[e.g.][]{Cheng:2023eba,Oayda:2024hnu}. Additionally, being ground-based, radio surveys are prone to declination-dependent sensitivity and flux calibration systematics.

The marginal significance of the cosmic dipole anomaly ended, however, with the advent of the catalogue of 1.4 million quasars produced by \citet{Secrest:2020has} using mid-infrared photometry from the CatWISE2020 \citep{2021ApJS..253....8M} data release of the Wide-field~Infrared~Survey~Explorer \citep[WISE;][]{Wright:2010qw}. \citet{Secrest:2020has} (hereafter \citetalias{Secrest:2020has}) showed that quasars exhibit a dipole amplitude over twice as large as expected from the CMB, although only $\sim\!28^\circ$ away in direction, and reported a $4.9\sigma$ discrepancy with the standard expectation, which was independently verified by \citet{Dam:2022wwh}. Moreover, with this dataset, the spectre of a significant clustering dipole appeared far less plausible since these quasars are estimated to have a mean redshift $\langle z \rangle = 1.2$, with 99\% having $z > 0.1$ \citepalias{Secrest:2020has}. Exploiting the near-total systematic independence of radio galaxies and mid-infrared quasars, a joint analysis by \citet{Secrest:2022uvx} (hereafter \citetalias{Secrest:2022uvx}) raised the significance with which the kinematic expectation is rejected to over $5\sigma$. Subsequently, a new catalogue from the Rapid ASKAP Continuum Survey, also in combination with the NVSS, has provided a $4.8\sigma$ confirmation of the dipole anomaly from radio data alone \citep{Wagenveld:2023kvi}. This was recently increased to over~$5\sigma$ by \cite{Bohme:2025nvu} with the addition of the LOFAR Two-Metre Sky Survey Data Release 2~\citep[LoTSS;][]
{Shimwell:2022oot}. The statistical significance of the cosmic dipole anomaly is no longer in question.

A lingering concern nevertheless has been whether unaccounted systematics in the catalogues or methods can give rise to a spurious dipole. For example, \citet{Abghari:2024eja} have argued that the significance of the CatWISE2020 quasar dipole anomaly reported by \citetalias{Secrest:2020has} has been overstated, based on their claim that the spatial distribution of the quasars may not be a pure dipole but contains low-order multipoles of comparable size. Since the quasar sample has about half of the sky symmetrically masked to mitigate a loss of completeness near the Galactic plane due to source confusion, the dominant quadrupole mode in the mask can couple the dipole with an octupole, potentially resulting in a biased estimate of the dipole amplitude that would make it less discrepant with the expectation from the CMB dipole. This is a testable hypothesis that simply requires appropriate methods for the extraction of low multipole signals on the sky~\citep[e.g.][]{Copi:2003kt,Hirata:2009ar,Gibelyou:2012ri,Oayda:2024voo}.

Investigation of power on  scales other than the dipole was previously undertaken by \citet{Tiwari:2022hnf}, who computed the angular power spectrum of the CatWISE2020 quasar catalogue. They confirmed the presence of an anomalously large dipole alongside the observation that the $\ell>30$ multipoles are concordant with the expectations of hierarchical structure formation as in $\Lambda$CDM. However, they also stated that \emph{all} multipoles with $\ell<10$ are anomalously high, despite their analysis not being quite suitable for drawing such a conclusion. This calls for a careful reassessment of the issue.

In this work, we revisit the clustering properties of the CatWISE2020 quasar sample. To this end, we present a consistent parameterisation of general low multipole templates that has a number of advantages (Section~\ref{sec:lowmultipoles}). We relate the parameters to angular power spectrum amplitudes, which allows us to assess their detectability given finite source counts (i.e.~shot noise). After discussing the catalogue's noise properties we define appropriate likelihoods from which we infer the low multipoles in various combinations, and compute Bayesian evidences for model comparison. We then turn to analysing  the higher multipoles in the catalogue's angular power spectrum (Section~\ref{sec:clustering}) which enables a data-driven estimation of the possible local clustering contribution to the observed quasar dipole. 

We confirm that the angular power spectrum of the quasar distribution on small scales is indeed consistent with the $\Lambda$CDM expectation. Howover, the largest scales ($2 \leq \ell \leq 4$) have power consistent with noise and have no significant impact on the dipole. This is paralleled in our real-space analysis of power in low multipoles  which, in particular, finds no evidence for a significant octupole. The dipole itself remains anomalously high and robust to all of the model variations we consider. In particular, it is not decreased by an octupole and is affected by less than $5\%$ (at $1\sigma$) given the random clustering dipole  predicted by the $\Lambda$CDM fit extrapolated to $\ell=1$.

\section{The CatWISE2020 Quasar Catalogue}
\label{sec:properties}

The CatWISE2020 quasar catalogue used in this work is the sample of mid-infrared quasars presented by \citetalias{Secrest:2022uvx}, selected via cuts on colour and magnitude from the full CatWISE2020 catalogue \citep{2021ApJS..253....8M}. The \citetalias{Secrest:2022uvx} catalogue differs from the version presented by \citetalias{Secrest:2020has} in being slightly deeper ($W1<16.5$ vs.\ $W1<16.4$) and using a refined sky mask. Both catalogues are affected by a source density trend with ecliptic latitude, first reported by \citetalias{Secrest:2020has}, which is likely due to the ecliptic being the principal axis of the WISE survey scanning pattern. The trend appears to be linear and is easily accounted for as a selection or weighting function. To remove poor-quality photometry near clumpy and resolved nebulae both in the Galaxy (e.g., planetary nebulae) and in nearby galaxies such as the Magellanic Clouds and Andromeda, a total of 48 specific sky regions were masked. Source confusion results in a drop off in source density at Galactic latitudes $|b| \leq 30^\circ$ given the depth of the sample, so this entire region is also masked.  This leaves a sample of 1,595,502 quasars.

Quasars are bolometrically dominant active galactic nuclei (AGN), which have an intrinsic luminosity exceeding that of the galaxy in which they reside, generated by the infalling matter in the accretion disk around their central supermassive black hole. Quasars are typically hosted in bluer, less massive, and less clustered galaxies~\citep{Hickox:2009ak} than radio AGN, and their continuum emission is mainly thermal and uncollimated (although their apparent luminosity at visual wavelengths is dependent on viewing angle because of their optically thick, dusty tori).  Higher abundance of gas-rich galaxies at redshifts $z\sim2-3$ drives increased accretion rates and quasar activity at that time, which, combined with preferential selection of high luminosity sources in large comoving volume, makes quasar samples moderate-to-high-redshift in general. Indeed the CatWISE2020 quasar catalogue of~\citetalias{Secrest:2022uvx} has a mean redshift $\langle z \rangle$ of 1.2 as shown via cross-matching with eBOSS redshifts (see Fig.~\ref{fig:redshiftdistribution} in Appendix~\ref{app:additional_figures}). This fulfils an  essential condition of the \citet{Ellis:1984uka} test, namely that sources should be at cosmological distances. Another requirement relates to the source spectra: the spectral energy distribution (SED) of quasars can have contributions from strong broadened emission features and from stellar populations of their host galaxies; however the large range of redshifts observed in quasars smears out steeply changing or discontinuous features, thus creating a smooth observed mean SED which can be sensibly modelled as a power-law: $S_\nu \propto \nu^{-\alpha}$~\citep[see, e.g., Figure~1 of][]{Secrest:2025wyu}. At the passband where the magnitude cut is performed, and when averaged over those sources with brightness close to this cut (the relevant average~\citep{vonHausegger:2024jan} for the Ellis \& Baldwin test), the sample of~\citetalias{Secrest:2022uvx} is well-characterised by spectral index $\alpha \simeq 1.06$. Measured at this cut, the power law index of the integral source counts is $x=1.71$, which determines the \citetalias{Ellis:1984uka} kinematic matter dipole expectation: $\mathcal{D}_{\rm kin}\approx0.73\times10^{-2}$. Lastly, the uniformity of the sample selection across the sky is influenced by the satellite scanning pattern along the ecliptic. \citetalias{Secrest:2020has} and \citetalias{Secrest:2022uvx} found the observed number density of their samples to decrease towards the ecliptic poles, exhibiting an approximately linear relation with ecliptic latitude. This trend vanishes with increasingly stringent flux cuts (see Fig.~\ref{fig:fluxcutvariation} below), suggesting that the cause is `Eddington bias', which allows sources with less certain flux measurements (such as those along the ecliptic equator where the satellite points less often) to up-scatter into the sample.  Regardless of the underlying mechanism, the selection with ecliptic latitude can simply be learned from the observed number density $\mathcal{N}_{\rm obs}$ on the sky by assuming an appropriate functional form for this trend, to then either define corresponding weights prior to fitting the data~(\citetalias{Secrest:2022uvx}), or to include it in an inference framework~\citep{Dam:2022wwh} as we too do here. In other words, the true signal, $\mathcal{S}=\mathcal{S}(\hat n)$, that underlies the number density of galaxies on the sky, is modulated as
\begin{align}
    \mathcal{N}_{\rm obs} = \mathcal{S}\left(1-f_{\rm ecl}\right) = \mathcal{S}\left(1-\frac{2}{\pi}|b_{\rm ecl}|\mathcal{Y}_{\rm ecl}\right),
    \label{eq:selection_ecl}
\end{align}
where $\mathcal{Y}_{\rm ecl}$ is the to-be-inferred amplitude of this linear trend with absolute ecliptic latitude, $|b_{\rm ecl}|$ (expressed in radians). In  this paper we formalise an inference for modulation of the sky by other effects as well. While the dipole modulation is of particular interest, we extend our inference to higher multipoles too, in order to detect any contamination or anomalous clustering on all scales.

\section{Low multipoles}
\label{sec:lowmultipoles}

Large-scale perturbations in the Universe hold significant implications for cosmology. The `low-$\ell$ anomalies' in the CMB~\citep{Schwarz:2015cma} raise questions also about their presence in the matter field~\citep[see][]{Schwarz:2015pqa}. While we await data from upcoming large-scale surveys, the CatWISE2020 quasar catalogue already offers insights in this regard. We focus on the lowest multipoles, specifically their potential influence on the anomalously large dipole.

Decomposing a signal on a masked sky into spherical harmonics requires methods that account for the breaking of their orthogonality on a $4\pi$ sky due to the mask. Some methods involve estimating the masks' coupling matrix (e.g. those related to `pseudo power spectrum' estimators, as used in the next section), while others evaluate the similarity of the signal with a given model in the unmasked part of the sky (as in the template-fitting we employ in this section). All methods of course need to be tested for possible biases. In the present context, this had been done by \citetalias{Secrest:2020has,Secrest:2022uvx} who showed that their least-squares estimator returned unbiased dipole amplitudes for Poisson-sampled data sets of sufficient size, in the absence of substantial power in higher multipoles. Although the sample size condition is well satisfied by the CatWISE2020 quasar catalogue, we must revisit this issue following the observation that the counts are overdispersed with respect to a Poisson distribution. We do so in this section, thereby  providing an update on the dipole measurement via Bayesian inference. In particular, we present a quantitative comparison of appropriate likelihoods and models, thus constraining the presence of higher multipoles and their influence on the dipole.

\subsection{Multipole amplitudes in real and in frequency space}\label{sec:lowmultipoles_translation}

We consider the number (density) of galaxies on the sky to be modulated by a combination of low multipoles as
\begin{align}
    \mathcal{N}_{\rm obs} = \mathcal{M}\left(1+f_\mathcal{D}+f_\mathcal{Q}+f_\mathcal{O}+\,...\right)\left(1-f_{\rm ecl}\right),\label{eq:numbercounts}
\end{align}
where $\mathcal{M}$ describes the unmodulated galaxy counts, considered here to be a constant monopole, and the terms $f$ describe modulating terms.  Among these is the linear ecliptic latitude correction, $f_{\rm ecl}$, Eq.~(\ref{eq:selection_ecl}), scaled by the free parameter $\mathcal{Y}_{\rm ecl}$, as well as templates ordered by spherical harmonic multipoles. We choose the following parametrisation for consistency with the usual dipole parametrisation:
\begin{align}
    f_\mathcal{D} =& \;\mathcal{D}\;\cos\alpha_\mathcal{D}\label{eq:fD}\\
    f_\mathcal{Q} =& \;\mathcal{Q}\left[\cos\alpha_\mathcal{Q}\cos\beta_\mathcal{Q}-\frac{1}{3}\cos\xi_{\alpha\beta}\right]\label{eq:fQ}\\
    f_\mathcal{O} =& \;\mathcal{O}\left[\cos\alpha_\mathcal{O}\cos\beta_\mathcal{O}\cos\gamma_\mathcal{O}\phantom{\frac{1}{5}}\right.\label{eq:fO}\\
    &\quad\left.-\frac{1}{5}\left(\cos\alpha_\mathcal{O}\cos\zeta_{\beta\gamma}+\cos\beta_\mathcal{O}\cos\zeta_{\gamma\alpha}+\cos\gamma_\mathcal{O}\cos\zeta_{\alpha\beta}\right)\right],\nonumber
\end{align}
describing general dipole, quadrupole, and octupole templates, respectively. This parameterisation presumes the definition of $\ell$ vectors per multipole $\ell$ (each defined by two angles on the sphere) that form $\ell-1$ angles $\alpha$, $\beta$, $...$ with the direction on the sky at which the signal $\mathcal{N}_{\rm obs}$ is evaluated, and that form deterministic angles $\xi,\zeta,...$ with each other.\footnote{E.g., a given direction $\hat n$ on the sky, forms angles with the two quadrupole vectors, denoted as $\hat p_{\mathcal{Q}1}$ and $\hat p_{\mathcal{Q}2}$, as $\cos\alpha_{\mathcal Q}=\hat p_{\mathcal{Q}1}\cdot\hat n$ and $\cos\beta_{\mathcal Q}=\hat p_{\mathcal{Q}2}\cdot\hat n$, while $\cos\xi_{\alpha\beta}=\hat p_{\mathcal{Q}1}\cdot\hat p_{\mathcal{Q}2}$ is constant for any one choice of the quadrupole vectors.  Equivalently any two of the three octupole vectors, e.g.~$\hat p_{\mathcal{O}1}$ and $\hat p_{\mathcal{O}2}$, give $\cos\alpha_{\mathcal O}=\hat p_{\mathcal{O}1}\cdot\hat n$ and $\cos\beta_{\mathcal Q}=\hat p_{\mathcal{O}2}\cdot\hat n$, while $\cos\zeta_{\alpha\beta}=\hat p_{\mathcal{O}1}\cdot\hat p_{\mathcal{O}2}$.}  Arbitrarily high multipole templates can be generated in this manner using methods dating back to \citet{Maxwell1954}.  This parameterisation also allows connection with known expressions of spherical harmonic functions. For instance, the $m=0$ mode of each of the terms $f$ is recovered by equating the multipole vectors for each $\ell$ such that $\alpha=\beta=...$ and $\xi=\zeta=...= 0$:
\begin{align}
    f_\mathcal{D}^{m=0} =& \;\mathcal{D}\;\cos\alpha_\mathcal{D},\\
    f_\mathcal{Q}^{m=0} =& \;\mathcal{Q}\left[\cos^2\alpha_\mathcal{Q}-\frac{1}{3}\right]\label{eq:fQ_m=0},\\
    f_\mathcal{O}^{m=0} =& \;\mathcal{O}\left[\cos^3\alpha_\mathcal{O}-\frac{3}{5}\cos\alpha_\mathcal{O}\right],
\end{align}
in a coordinate system where the multipole vectors are aligned with the spherical harmonics' zenith.  To enable connection between the findings of this and the following section, we first discuss the relation between the power spectrum amplitude $C_\ell$ and the amplitude $\mathcal{D}$, $\mathcal{Q}$, $\mathcal{O}$, etc of the multipole templates.

It is well known that the amplitudes of the monopole and dipole, $\mathcal{M}$ and $\mathcal{D}$, as defined in Eq.~(\ref{eq:numbercounts}) and following, are related to the  corresponding $C_\ell$ via $C_0 = 4\pi\mathcal{M}^2$ and $C_1 = 4\pi\mathcal{M}^2\mathcal{D}^2/9$. However, since we usually compute the power spectrum from an overdensity map, i.e.~one that is normalised by its monopole, we obtain instead:
\begin{align}
    C_0 &= 4\pi\label{eq:C0},\\
    C_1 &= \frac{4\pi}{9}\mathcal{D}^2\label{eq:C1}.
\end{align}
For higher multipoles $\ell>1$ there is no such one-to-one relation, since the different combinations of angles in Eqs.~(\ref{eq:fQ}),~(\ref{eq:fO}), etc. at fixed multipole amplitude, correspond to different spherical harmonic coefficients $a_{\ell m}$, and so generally combine in ways that span a range in overall average power $C_\ell$.  Indeed, expressing the higher order multipoles with one common amplitude each distinguishes this particular parametrisation from others that may simply consider each spherical harmonic template to be scaled with its own amplitude.  Nevertheless, to translate between an observed power spectrum and corresponding multipole amplitudes, it is helpful to compute the range of power $C_\ell$ corresponding to a given multipole amplitude and vice versa.  Here we make use of the fact that the spherical harmonics are orthogonal functions on the sphere and the power spectrum is computed over the average of their squared amplitudes, hence the identification of any one mode as giving rise to the smallest or largest power at that multipole will remain valid even when considering a sum of modes. The mode for which a given multipole amplitude has the largest relative contribution to the power spectrum is always the $m=0$ mode, so the maximum power can be read off the corresponding spherical harmonic's $Y_{\ell 0}$ coefficient. For the quadrupole and octupole, these maximum values read:\footnote{Using $Y_{\ell 0}=\sqrt{\frac{2\ell+1}{4\pi}}P_\ell$ and expressing the Legendre polynomials $P_\ell$ in terms of generalized binomial coefficients, one finds the general relation: ${\rm max}(C_\ell)=\frac{4\pi}{2^{2\ell}(2\ell+1)^2}\binom{\ell-0.5}{\ell}^{-2}\,\mathcal{A}_\ell^2$ for a given multipole's amplitude $\mathcal{A}_\ell$.}
\begin{align}
    {\rm max}(C_2) &= \frac{16\pi}{9\cdot25}\mathcal{Q}^2,\label{eq:C2max}\\
    {\rm max}(C_3) &= \frac{16\pi}{25\cdot49}\mathcal{O}^2.\label{eq:C3max}
\end{align}
By inspection, the smallest contribution is from either of the $m=\pm1, \pm2$ modes for the quadrupole and the $m=\pm2$ modes for the octupole. These minimum values read:
\begin{align}
    {\rm min}(C_2) &= \frac{4\pi}{3\cdot25}\mathcal{Q}^2\label{eq:C2min}\\
    {\rm min}(C_3) &= \frac{4\pi}{3\cdot5\cdot49}\mathcal{O}^2\label{eq:C3min}
\end{align}
Conversely, the range of multipole amplitudes allowed by a fixed power $C_\ell$ are:
\begin{align}
    \mathcal{Q} &\in \left[\frac{3\cdot5}{4}\sqrt\frac{C_2}{\pi},\frac{5}{2}\sqrt\frac{3\cdot C_2}{\pi}\;\right],\label{eq:Qrange}\\
    \mathcal{O} &\in \left[\frac{5\cdot7}{4}\sqrt\frac{C_3}{\pi},\frac{7}{2}\sqrt\frac{3\cdot 5\cdot C_3}{\pi}\;\right].\label{eq:Orange}
\end{align}
These considerations are helpful when comparing multipole amplitudes with power spectrum amplitudes.  For example, considering a signal on the sky for which dipole and octupole have the same power, $C_3=C_1$, we can see from Eqs.~(\ref{eq:C1}) and~(\ref{eq:Orange}) that $\mathcal{O}\in\left[\mathcal{D}\cdot35/6,\mathcal{D}\cdot7\sqrt{5/3}\right]$, i.e. the \emph{same} power corresponds to a much larger value of the octupole amplitude. 

\subsection{``Shot noise multipoles''}\label{sec:lowmultipoles_shotnoise}

We can now estimate the detectability of each multipole given finite source counts. $N$ sources, randomly sampled from an unmodulated sky, \mbox{$\mathcal{N}_{\rm obs}=\mathcal{M}$}, contain random, uncorrelated fluctuations on all scales, i.e. shot noise which contributes a variance per multipole of $N_\ell=4\pi/N$. Hence for a given number of sources $N$, a particular multipole would be considered detectable only if its amplitude is larger than this noise.

Similarly to earlier estimations of the `shot noise dipole'~\citep{Baleisis:1997wx,Secrest:2025akp} which were made by equating $N_\ell$ to $C_1$ from Eq.~(\ref{eq:C1}), we can now compare any multipole amplitude with the shot noise expected for a given number $N$ of sources, using the relations above. As examples, we compute the `shot noise quadrupoles' and `shot noise octupole' in the ranges just discussed:
\begin{align}
    \mathcal{Q}_{\rm SN} &\in \left[\frac{3\cdot 5}{2\sqrt{N}},\frac{5\sqrt{3}}{\sqrt{N}}\right],\label{eq:QSN}\\
    \mathcal{O}_{\rm SN} &\in \left[\frac{5\cdot 7}{2\sqrt{N}},\frac{7\sqrt{3\cdot5}}{\sqrt{N}}\right].\label{eq:OSN}
\end{align}
Optimistically considering the lower limits of relations (\ref{eq:QSN}) and (\ref{eq:OSN}), to be able to detect a quadrupole of amplitude $\mathcal{Q}$ at $3$ times the noise level $\mathcal{Q}_{\rm SN}$ thus requires $N\gtrsim (22.5
/\mathcal{Q})^2$ sources distributed over the full sky, and similarly $N\gtrsim (52.5/\mathcal{O})^2$ for an octupole of amplitude $\mathcal{O}$.  Taking into account the sky fraction in which sources are actually observed:
\begin{align}
    \mathcal{Q}&\gtrsim3\cdot\mathcal{Q}_{\rm SN}=22.5\,\sqrt{f_{\rm sky}/N},\label{eq:QSNdet}\\
    \mathcal{O}&\gtrsim3\cdot\mathcal{O}_{\rm SN}=52.5\,\sqrt{f_{\rm sky}/N}.\label{eq:OSNdet}
\end{align}
For our CatWISE2020 quasar catalogue~(\citetalias{Secrest:2022uvx}) where \mbox{$N\approx 1.6\times10^6$} and $f_{\rm sky}\approx0.5$, this means that a quadrupole with amplitude $\mathcal{Q}\gtrsim0.013$ is detectable. However, an octupole would have to be as strong as $\mathcal{O}\gtrsim0.029$ in order to be detectable at $3$ times the shot noise. We quantify this in more detail below.

Considering the effect Poisson shot noise has on the measurement of a dipole also informed the frequentist simulations of~\citetalias{Secrest:2020has} and~\citetalias{Secrest:2020has}, with which they quantified the $p$-value of obtaining the measured number count dipole given the expected kinematic dipole. It has been noted subsequently~\citep[e.g.][]{Abghari:2024eja} that the possible presence of higher multipoles in the data, in particular of an octupole, ought to be reflected in such simulations as well, as this can influence the claimed significance. The computations in this section will place such considerations in context. We provide a detailed rebuttal of \citet{Abghari:2024eja} in Appendix~\ref{app:lowmultipoles_correlation}.

Our primary aim, however, is to formalise an inference that respects the true noise properties of the data, in order to provide a robust measurement of the dipole amplitude and simultaneously quantify the evidence for or against the presence of higher-multipole signals. This inference must  account for correlations among the multipoles in the presence of a mask. This is done in the next sections where we trade the frequentist framework and Poisson simulations for a fully Bayesian inference framework using non-Poissonian likelihoods.

\subsection{Super-Poisson clustering and one-point statistics}
\label{sec:lowmultipoles_onepoint}

Counts-in-cells statistics of extragalactic sources are often modelled by drawing random point samples from a perturbed density field, where each cell's assigned density determines the probability of a source being sampled within it, e.g.~by scaling the rate parameter of a Poisson process. In this example, one finds the underlying distribution of the source count in a given cell to be Poissonian; the distribution across all cells, however, will in general \textit{not} be Poissonian, as the rate parameter itself would have been drawn from a prior distribution (the distribution of density values across cells). The net result is a `Mixed Poisson distribution'~\citep[e.g.][]{Karlis:2005abc}.\footnote{Consider the Poisson distribution,
\begin{align}
    P(X=k) = \frac{1}{k!}\lambda^k\exp(-\lambda),
\end{align}
with rate parameter $\lambda$ sampled from a prior probability distribution, $\Pi$. If the prior is a Gamma distribution $\Pi(\lambda) \sim \Gamma(r,s)$, with `concentration' $r$ and `shape' $s=(1-p)/p$, the integration over the prior leads to the `Gamma-Poisson distribution',
\begin{align}
    P(X=k) = \frac{\Gamma(k+r)}{\Gamma(r)\,k!}\,(1-p)^k\,p^r,
\end{align}
which is an example of a Mixed Poisson distribution -- in this case a `super-Poissonian' distribution due to its additional dispersion.  The Gamma-Poisson distribution is often referred to as `negative binomial distribution' although we will prefer the former here.}
Moreover, at fixed cell size, a number of effects generally present in source catalogues have been shown to alter the one-point statistics of a given cell. For instance, radio galaxies exhibit radio lobes that can be misidentified as individual sources, thus introducing small-scale spatial correlations, as can imaging artefacts. The one-point distributions over all cells' source counts 
do not distinguish between these effects -- radio lobes/imaging artefacts on small scales, and cosmological perturbations on large scales -- so one must resort to two-point statistics which allow for appropriate scale separation, as was done by~\citet{Blake:2004dg} using the NVSS sample.

Nevertheless, regardless of their physical origin, it is essential to model the observed one-point distributions with suitable functions to ensure accurate inference of signals underlying the counts-in-cells. Studies of super-Poissonian counts-in-cells in projected fields were performed recently by, e.g., \citet{Siewert:2019poc} and \citet{Pashapour-Ahmadabadi:2025ghj} using the LoTSS-DR1 and -DR2 data, respectively. There have been  similar analyses of the statistics of gravitational clustering~\citep[e.g.][]{1984ApJ...276...13S,1995MNRAS.274..213S,Kitaura:2013cwa} focussing on two such distributions, the `compound-Poisson' and `negative binomial' distribution.

\begin{figure}
    \centering
    \includegraphics[width=\columnwidth]{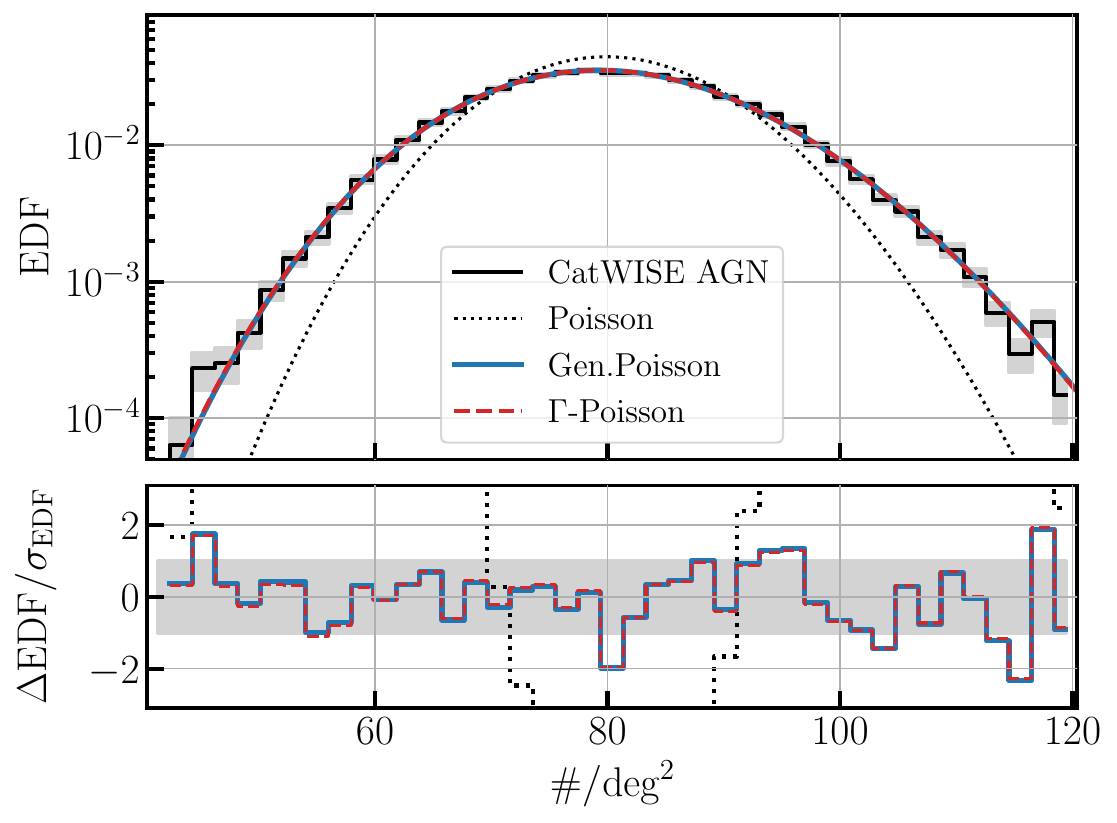}
    \caption{Empirical one-point distribution function (EDF) of the CatWISE2020 quasar number densities (black) after correction for the ecliptic latitude trend (\ref{eq:selection_ecl}) and removal of the best-fit dipole. \textit{Top panel:} Comparison of the EDF against a Poisson (dotted black), Generalised Poisson (blue), and Gamma Poisson (=\,negative-binomial) distribution (dashed red).  For visual clarity these are shown as continuous distributions.  The grey-shaded error band indicates Poisson errors per bin. \textit{Bottom panel:} Difference of the respective distributions from the EDF normalised to the error per bin, indicating that both the super-Poissonian distributions provide excellent fits to the data, as opposed to the strongly deviant Poisson distribution.}
    \label{fig:1pt}
\end{figure}

This is especially relevant for the choice of dipole estimator, or, correspondingly, a suitable likelihood. Note that \citet{Dam:2022wwh} and  \citet{Wagenveld:2025ewl} defined a Poisson likelihood with which they inferred the matter dipole in the CatWISE2020 samples of \citetalias{Secrest:2020has} and \citetalias{Secrest:2022uvx}, respectively. However as seen in Fig.~\ref{fig:1pt}, the distribution over all pixels' source densities (black) clearly demonstrates overdispersion with respect to a Poisson distribution (dotted black) whose mean value has been fixed to that of the observed distribution. It turns out that this overdispersion is indeed well described by a negative binomial distribution (dashed red), although it remains to be demonstrated how the generation of such a distribution is related to the underlying mechanism that gives rise to observed source counts in a given catalogue \citep[cf.][]{Pashapour-Ahmadabadi:2025ghj}.  Accordingly we will refer to it more generally as Gamma-Poisson distribution (see footnote 4).  For the same reason we consider an alternative, phenomenological distribution, the `generalised Poisson distribution'~\citep[see, e.g.][]{1984ApJ...276...13S} whose free parameter scales the variance of the resulting distribution, while keeping the mean unchanged with respect to that of a regular Poisson (blue); this describes the counts equally well as seen in Fig.~\ref{fig:1pt}.  (We introduce expressions for all these distributions in Section~\ref{sec:lowmultipoles_inference}.)

It is thus evident that a pure Poisson likelihood does not describe the data well and inferring the matter dipole with either of these super-Poissonian likelihoods is more appropriate. This was done recently by~\citet{Bohme:2025nvu} who re-ran inferences of a number of radio samples with a negative binomial likelihood to arrive at a \emph{combined} statistical significance of $>5\sigma$ for the matter dipole anomaly from radio data alone. In the next section, we too explore super-Poissonian likelihoods with the CatWISE2020 quasar catalogue, still the largest single data set that establishes the dipole anomaly at equally high significance. In addition, we perform inferences of higher multipoles.

\subsection{Inference procedure}
\label{sec:lowmultipoles_inference}

We jointly infer the monopole, dipole, and higher multipole modes in various combinations of the $N_{\rm side}=64$ CatWISE2020 quasar number density map of \citetalias{Secrest:2022uvx} using a selection of relevant likelihoods.  To this effect, we utilize the `No U-Turn' Hamiltonian Monte Carlo sampler~\citep{Hoffman:2011ukg} of \textsc{numpyro}~\citep{Phan:2019elc}, and in each run create sufficiently large samples to satisfy the \citet{Gelman:1992zz} convergence statistic $|\widehat{R}-1|<10^{-2}$.  Finally, we compute the Bayesian evidence for each model via the harmonic approximation using \textsc{harmonic}~\citep{Polanska:2024arc}.

In light of the discussion in Section~\ref{sec:lowmultipoles_onepoint} we consider various likelihoods to describe the data.  While we have established already that the data is not described well by a Poisson process with a rate parameter $\lambda$ that equates solely to the modulated sky model, we still consider as a starting point, and to make connection with previous work \citep[e.g.][]{Dam:2022wwh,Wagenveld:2023kvi}, a Poisson likelihood per pixel:
\begin{align}
    \ln \mathcal{L}(N_{\rm obs}|\lambda) = N_{\rm obs}\ln\lambda - \sum_{n=1}^{N_{\rm obs}}\ln n \,-\lambda, \label{eq:lnl_poisson}
\end{align}
where $\lambda=\mathcal{N}_{\rm obs}$ is the modelled rate parameter in a given pixel~(Eq.~\ref{eq:numbercounts}), and $N_{\rm obs}$ is the number of observed sources in that pixel.

While the 1-point distribution over all pixels of the map agrees well with either of the super-Poissonian distributions considered in Section~\ref{sec:lowmultipoles_onepoint}, we do not assume this \emph{a priori} to also hold for the counts in individual pixels scattered around $\mathcal{N}_{\rm obs}$. So we consider these distributions alongside one another, to be  eventually compared according to their Bayesian evidences. The Generalised Poisson likelihood,
\begin{align}
    \ln \mathcal{L}(N_{\rm obs}|\lambda,b) =& \,\ln\left(\lambda(1-b)\right) + (N_{\rm obs}-1)\ln\left(\lambda(1-b)+N_{\rm obs}b\right) \nonumber\\
    &- \sum_{n=1}^{N_{\rm obs}}\ln n \,- \lambda(1-b) - N_{\rm obs}b,\label{eq:lnl_genpoisson}
\end{align}
where $\lambda=\mathcal{N}_{\rm obs}$ is the modelled rate parameter in a given pixel as before, also takes the global, free parameter $b$ to model the observed counts in that pixel.  The Gamma-Poisson likelihood reads:
\begin{align}
    \ln \mathcal{L}(N_{\rm obs}|r,p) =& \,\ln \Gamma (N_{\rm obs}+r) - \ln \Gamma(r)\nonumber\\
    &- \sum_{n=1}^{N_{\rm obs}}\ln n \,+ N_{\rm obs}\ln(1-p) + r\ln(p),\label{eq:lnl_gammapoisson}
\end{align}
where $r=\mathcal{N}_{\rm obs}p/(1-p)$ is the modelled concentration parameter in each pixel, $\Gamma$ is the gamma function, and here the global, free parameter $p$ models the counts' dispersion.  In principle, mixed Poisson models of various kinds could be explored here, but given the excellent agreement between the Generalized Poisson and the Gamma-Poisson models reported below, we deem our present choice adequate for the description of the data at hand.

\begin{table}
    \centering
    \begin{tabular}{c|l|l}
     Model 	& Alias 					& Free parameters \\ \hline
     (i) 	& ${\sf D}$ 				& $\mathcal{M}$, $\mathcal{D}$, +2 angles \\
     (ii) 	& ${\sf DY}_{\sf ecl}$ 		& $\mathcal{M}$,$\mathcal{D}$, $\mathcal{Y}_{\rm ecl}$, +2 angles \\
     (iii) 	& ${\sf DQ}$ 				& $\mathcal{M}$, $\mathcal{D}$, $\mathcal{Q}$, +6 angles \\
     (iv) 	& ${\sf DY}_{\sf ecl}{\sf Q}$ 	& $\mathcal{M}$, $\mathcal{D}$, $\mathcal{Q}$, $\mathcal{Y}_{\rm ecl}$, +6 angles \\
     (v) 	& ${\sf DY}_{\sf ecl}{\sf O}$ 	& $\mathcal{M}$, $\mathcal{D}$, $\mathcal{O}$, $\mathcal{Y}_{\rm ecl}$, +8 angles \\
     (vi)   	& ${\sf DQO}$ 				& $\mathcal{M}$, $\mathcal{D}$, $\mathcal{Q}$, $\mathcal{O}$, +12 angles \\
     (vii)  	& ${\sf DY}_{\sf ecl}{\sf QO}$ 	& $\mathcal{M}$, $\mathcal{D}$, $\mathcal{Q}$, $\mathcal{O}$, $\mathcal{Y}_{\rm ecl}$, +12 angles \\
     (viii)   	& ${\sf DQOH}$			& $\mathcal{M}$, $\mathcal{D}$, $\mathcal{Q}$, $\mathcal{O}$, $\mathcal{H}$, +20 angles \\
    \end{tabular}
    \caption{List of models and their free parameters. Depending on the choice of likelihood (Eqs.~(\ref{eq:lnl_poisson}), (\ref{eq:lnl_genpoisson}) and (\ref{eq:lnl_gammapoisson})), an additional free parameter from $\{p, b\}$ may be included.  Each included general multipole $\ell$ entails $\ell$ multipole vectors each with two free angles.}
    \label{tab:models}
\end{table}

For each likelihood, we consider various models of increasing richness to describe the data, based on the terms included in Eq.~(\ref{eq:numbercounts}).  Table~\ref{tab:models} lists the considered combinations of free parameters, where those not listed are fixed to zero, respectively.  If sampled, the parameters' priors were chosen to be:
\begin{align}
    \mathcal{M} &\sim \mathcal{U}(0,200)\\
    \mathcal{D},\mathcal{Q},\mathcal{O},\mathcal{H} &\sim \mathcal{U}(0,1)\\
    \mathcal{Y}_{\rm ecl} &\sim \mathcal{U}(-2,2)\\
    \noalign{\smallskip} \hline \noalign{\smallskip}
    \phi &\sim \mathcal{U}(0,2\pi)\\
    \cos\theta &\sim \mathcal{U}(-1,1)\\
    \noalign{\smallskip} \hline \noalign{\smallskip}
    p,b &\sim \mathcal{U}(0,1)
\end{align}
When fitting all multipole angles of either quadrupole and octupole we further restrict the angles' priors such that $\phi_{\mathcal{Q}2}\leq\phi_{\mathcal{Q}1}$ and $\phi_{\mathcal{O}1}\leq\phi_{\mathcal{O}2}\leq\phi_{\mathcal{O}3}$~\citep[as, e.g., in ][]{Patel:2024btm} although this does not make any appreciable difference.

\subsection{Inference results}
\label{sec:lowmultipoles_nooctupole}

\subsubsection{The ecliptic latitude trend}

The linear ecliptic latitude trend present in both the \citetalias{Secrest:2020has} and \citetalias{Secrest:2022uvx} samples was initially corrected in the data prior to applying the least-squares dipole estimator.  To account for the uncertainty in the trend, the simulations, with which the significance was estimated, were later modulated by this trend within the fit uncertainties in \citetalias{Secrest:2022uvx}. The re-analysis by \citet{Dam:2022wwh} finally integrated the trend into a simultaneous Bayesian inference of monopole, dipole, and ecliptic latitude trend, such that the data could be analysed without prior correction, and with the advantage that correlations between the free parameters would be consistently retained. Their results, using the data of \citetalias{Secrest:2020has}, confirmed our finding of the anomalously large dipole\footnote{That the symmetric, even-$\ell$ ecliptic latitude trend does not influence the recovered dipole amplitude is demonstrated in Appendix~\ref{app:clustering_ecliptic}.} and, by computing Bayes factors, showed that a model that includes only a free monopole and dipole (model (i) in our Table~\ref{tab:models}) is outscored by one that also includes the ecliptic latitude (model (ii) in our Table~\ref{tab:models}), underscoring its importance in describing the data correctly.  We revisit this inference and evidence computation and show in the upper section of Table~\ref{tab:evidenceratios} that a Poisson likelihood indeed returns overwhelming evidence for model (ii) over model (i), just as found by \citet{Dam:2022wwh}. Here $\Delta\log_{10}(\mathcal{Z}) = \log_{10}(\mathcal{Z_{\rm (ii)}/\mathcal{Z}_{\rm (i)}})$ and $\mathcal{Z}_{(a)}$ is the Bayesian evidence of model $(a)$.  The difference in the two Poisson evidences arises simply due to our using the updated data set of \citetalias{Secrest:2022uvx} here, while \citet{Dam:2022wwh} used that of \citetalias{Secrest:2020has}.

\begin{table}
    \centering
    \begin{tabular}{l|l|l}
    Model comparison & Likelihood & $\Delta\log_{10} \mathcal Z$ \\\hline
    ${\sf DY}_{\sf ecl}$ vs. {\sf D}      
    &   Poisson                     & 140.09\\
    &   Poisson \citep{Dam:2022wwh} & 78.18\\
    &   Generalised Poisson         & 106.99\\
    &   Gamma Poisson               & 106.98\\
    \hline
    {\sf DQ} vs. {\sf D}      
    &   Generalised Poisson         & 106.41\\
    {\sf DQ} vs. ${\sf DY}_{\sf ecl}$      
    &   Generalised Poisson         & -0.58\\
    ${\sf DY}_{\sf ecl}{\sf Q}$ vs. ${\sf DY}_{\sf ecl}$
    &   Generalised Poisson         & 1.75\\
    ${\sf DY}_{\sf ecl}{\sf Q}$ vs. {\sf DQ}
    &   Generalised Poisson         & 2.33\\
    \hline
    ${\sf DY}_{\sf ecl}{\sf O}$ vs. ${\sf DY}_{\sf ecl}$      
    &   Generalised Poisson         & -0.79\\
    {\sf DQO} vs. ${\sf DQ}$      
    &   Generalised Poisson         & -0.81\\
    {\sf DQO} vs. ${\sf DY}_{\sf ecl}$      
    &   Generalised Poisson         & -1.35\\
    ${\sf DY}_{\sf ecl}{\sf QO}$ vs. ${\sf DY}_{\sf ecl}{\sf Q}$
    &   Generalised Poisson         & -0.81\\
    \hline
    \end{tabular}
    \caption{Evidence ratios for any two models in the left column, computed for the CatWISE2020 quasar sample of \citetalias{Secrest:2022uvx} and using likelihoods as listed in the middle column. $\Delta\log_{10}\mathcal{Z}$ is the difference between Bayes factors of the first- and the second-mentioned model in the left column; a positive value corresponds to the preference of the first-mentioned model. The evidence of \citet{Dam:2022wwh} was computed using the sample of \citetalias{Secrest:2020has}.  No evidence is found for models including an octupole.}
    \label{tab:evidenceratios}
\end{table}

\subsubsection{Preference for super-Poissonian likelihoods}

We also run the inference on the same two models using the Generalised Poisson (`GenPo') and the Gamma Poisson likelihood, Eqs.~(\ref{eq:lnl_genpoisson}) and~(\ref{eq:lnl_gammapoisson}).  Also in the top section of Table~\ref{tab:evidenceratios} it is seen that the overwhelming preference for model (ii) over (i) is given for these likelihoods as well, where the slightly lower Bayes factor when compared against the Poisson (`Po') likelihood stems from the fact that an extra parameter each governs the over-dispersion and so is penalised.  Given a fixed model, however, the Bayes factor is clearly in favour of the super-Poissonian likelihoods.  For instance $\log_{10}(\mathcal{Z}^{\rm GenPo}_{\rm (ii)}/\mathcal{Z}^{\rm Po}_{\rm (ii)})=190.38$ for model (ii).  This was indeed expected given our observations in the one-point distributions of the best-fit removed data in Section~\ref{sec:lowmultipoles_onepoint} and Fig.~\ref{fig:1pt}.

\begin{figure}
    \includegraphics[width=\columnwidth]{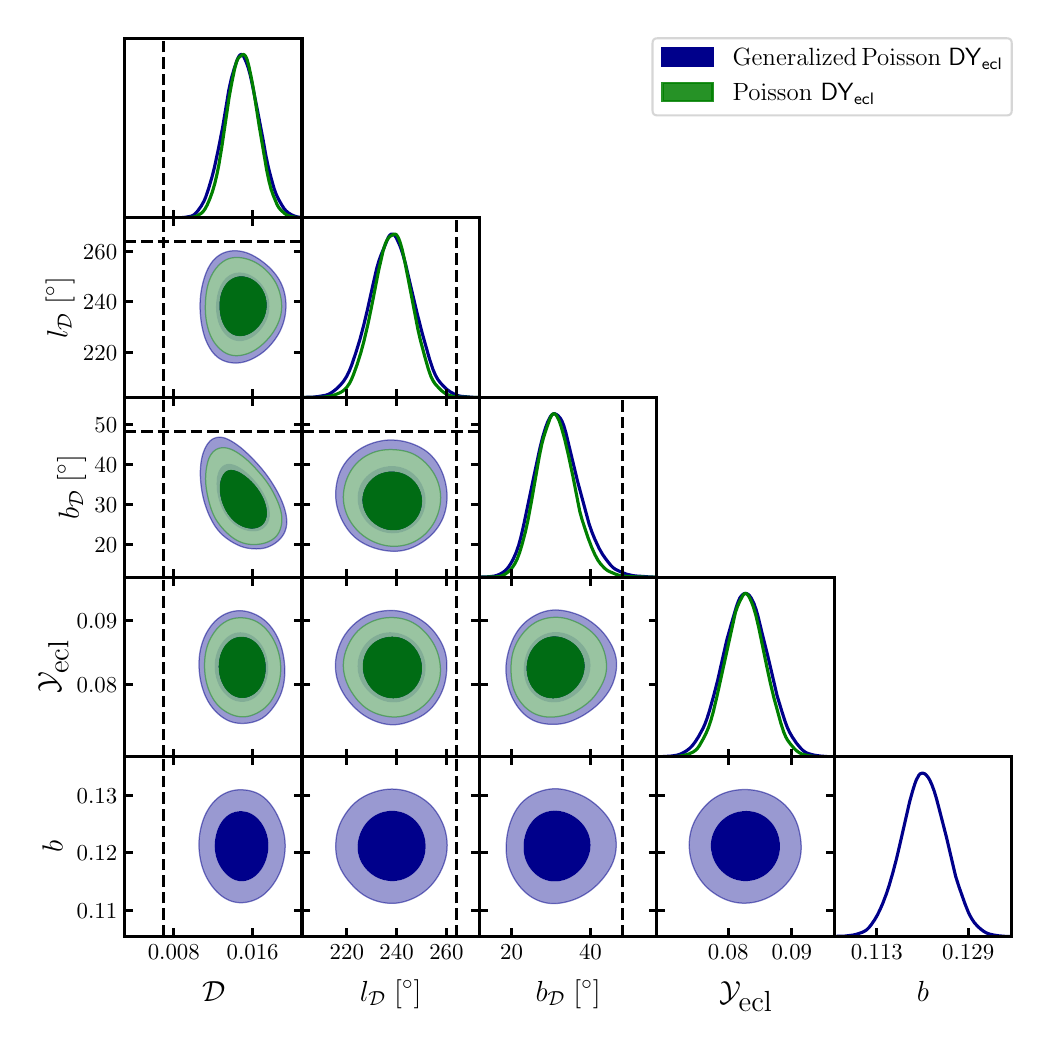}
    \caption{Posteriors of all parameters in model (ii) (${\sf DY}_{\sf ecl}$, see Table~\ref{tab:models}) inferred using the Poisson and the Generalised Poisson likelihood, Eqs.~(\ref{eq:lnl_poisson}) and~(\ref{eq:lnl_genpoisson}).  The dashed lines indicate the expected amplitude and direction for the kinematic dipole $\mathcal{D}_{\rm kin}$.}
    \label{fig:decl_like_comparison}
\end{figure}

The posteriors of model (ii) computed with the Poisson and Generalised Poisson likelihoods are shown in Fig.~\ref{fig:decl_like_comparison}.  Apart from the additional parameter $b$ of the Generalized Poisson likelihood that is shown in the bottom row, the posteriors appear largely concentric and exhibit virtually no change in the parameters' mean values, although their widths are slightly enlarged in the case of the Generalised Poisson likelihood relative to the purely Poissonian, consistent with the observations of \citet{Bohme:2025nvu} using radio data. In particular, the dipole amplitude in this model is inferred to be
\begin{align}
&\hat{\mathcal{D}}^{\rm Po}_{\rm (ii)}=0.0149_{-0.0016}^{+0.0016} \ , && \hat{\mathcal{Y}}^{\rm Po}_{\rm ecl,(ii)} = 0.0827_{-0.0032}^{+0.0032} \ ,\label{eq:po_decl_dipoleamplitude}\\
&\hat{\mathcal{D}}^{\rm GenPo}_{\rm(ii)}=0.0149_{-0.0018}^{+0.0018} \ ,&& \hat{\mathcal{Y}}^{\rm GenPo}_{\rm ecl,(ii)} = 0.0827_{-0.0036}^{+0.0036} \ ,
\label{eq:genpo_decl_dipoleamplitude}
\end{align}
 assuming a Poisson and the Generalised Poisson likelihood, respectively. The posteriors obtained with the Gamma Poisson likelihood are virtually the same as those from the Generalised Poisson likelihood; in Table~\ref{tab:evidenceratios2} we present evidence ratios for both models over the Poissonian. It is seen that for our choice of priors both models are nearly equally preferred by the data. While the Gamma Poisson likelihood appears slightly favoured with respect to the Generalised Poisson likelihood (for all models considered), the latter allows for faster sampling in our implementation. Hence we  present below only the results with the Generalised Poisson likelihood.

\begin{table}
    \centering
    \begin{tabular}{l|l|l}
    Model & Likelihood comparison & $\Delta\log_{10} \mathcal Z$ \\\hline
    ${\sf D}$
    &   Generalised Poisson vs.~Poisson         & 223.48\\
    &   Gamma Poisson vs.~Poisson               & 223.68\\
    &   Generalised vs.~Gamma Poisson           & -0.204\\
    \hline
    ${\sf DY}_{\sf ecl}$
    &   Generalised Poisson vs.~Poisson         & 190.38\\
    &   Gamma Poisson vs.~Poisson               & 190.58\\
    &   Generalised vs.~Gamma Poisson           & -0.200\\
    \hline
    \end{tabular}
    \caption{Evidence ratios for the models in the left column, computed for the CatWISE2020 quasar sample of \citetalias{Secrest:2022uvx}, and comparing any two likelihoods as listed in the middle column.  $\Delta\log_{10}\mathcal{Z}$ is the difference between Bayes factors of the first- and the second-mentioned likelihood in the middle column; a positive value corresponds to the preference of the first-mentioned model.}
    \label{tab:evidenceratios2}
\end{table}

\subsubsection{Quadrupole and ecliptic latitude trend}

The ecliptic latitude trend can be described to first order by a $m=0$ quadrupole in a coordinate system aligned with the ecliptic, cf.~Eq.~(\ref{eq:fQ_m=0}).  However, one ought to allow also for other quadrupolar modes in the data, so in model (iii) we consider a general quadrupole in addition to monopole and dipole.  
\begin{align}
    \hat{\mathcal{D}}^{\rm GenPo}_{\rm (iii)} = 0.0151_{-0.0018}^{+0.0019} ,&&
    \hat{\mathcal{Q}}^{\rm GenPo}_{\rm (iii)} =
    0.0784_{-0.0040}^{+0.0041} .
\end{align}
We compare the marginalised likelihood of this model with that of model (ii) which, instead of a general quadrupole, contained only the linear ecliptic latitude correction.  The middle section of Table~\ref{tab:evidenceratios} shows that the data clearly prefers a quadrupolar correction over no correction of the ecliptic latitude trend, although a general quadrupolar correction appears less favoured than the linear ecliptic modulation by $\log_{10}(\mathcal{Z}_{\rm (iii)}/\mathcal{Z}_{\rm (ii)})=-0.58$, which is ``substantial'' evidence according to the scale of \citet{Kass:1995loi}.  This is possibly due to the increased number of parameters of model (iii) compared with model (ii), given that the recovered quadrupole angles are constrained to be close to the ecliptic axis, which is simply fixed in model (ii).

We further consider in model (iv) a general quadrupole in addition to the linear ecliptic latitude trend. As also seen in the middle section of Table~\ref{tab:evidenceratios}, this model is indeed favoured over models (ii) and (iii) with ``strong'' to ``decisive'' evidence, owing to the appearance of a non-zero quadrupole amplitude in addition to a nearly unchanged dipole amplitude:
\begin{align}
    \hat{\mathcal{D}}^{\rm GenPo}_{\rm (iv)} = 0.0148_{-0.0018}^{+0.0018} ,\nonumber\\
    \hat{\mathcal{Q}}^{\rm GenPo}_{\rm (iv)} = 0.0241_{-0.0061}^{+0.0084} ,\\
    \hat{\mathcal{Y}}^{\rm GenPo}_{\rm ecl,(iv)} = 0.0893_{-0.0156}^{+0.0194} .\nonumber
\end{align}
We note that the recovered posteriors on the quadrupole angles, while rather extended, are consistent with being located at the ecliptic poles, wherefore this finding indicates the necessity of a correction to the assumed perfectly linear shape of the ecliptic latitude trend.  While the numerical value for the median $\mathcal{Q}$ is larger than that of $\mathcal{D}$, the association of the quadrupole with the ecliptic latitude trend, whose amplitude $\mathcal{Y}_{\rm ecl}$ already is of this order, prevents a clear interpretation of the found quadrupole as an independent detection of higher multipole power in the data.  However, regardless of its interpretation, we find the dipole amplitude to remain consistent with those values found in the preceding models. We present select posteriors of models (ii), (iii) and (iv), in Fig.~\ref{fig:declq_triangle} of Appendix~\ref{app:additional_figures} and return to the question about angular power in the quadrupole in Section~\ref{sec:clustering_powerspectrum}.

\subsubsection{No evidence for an octupole in CatWISE2020 quasars}

We finally turn to inferences including general octupoles to quantify any correlation that might exist with the dipole, which might have altered its best-fit value, and to quantify the evidence for an octupole's presence in the data altogether.  To this effect, we consider two models, model (v) where a general octupole is added to model (ii), including the linear ecliptic latitude modulation, and model (vi) where instead all multipoles $\ell\leq3$ are included in the inference.  

\begin{figure}
    \centering
    \includegraphics[width=\columnwidth]{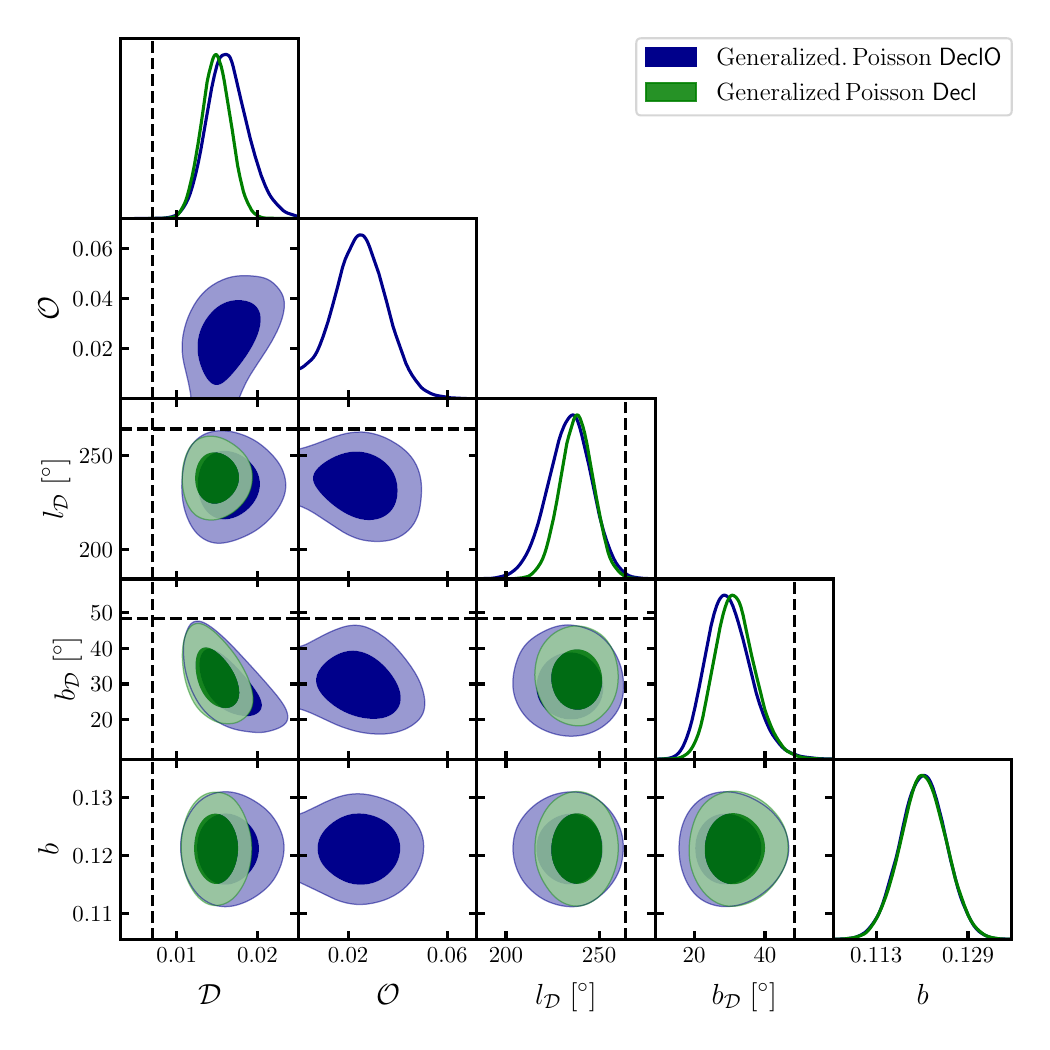}
    \caption{Posteriors of select parameters in models (v) and (ii) (${\sf DY}_{\sf ecl}{\sf O}$ and ${\sf DY}_{\sf ecl}$, see Table~\ref{tab:models}) inferred using the Generalised Poisson likelihood (Eq.~(\ref{eq:lnl_genpoisson})).  The posteriors are marginalised over the octupole angles and the ecliptic latitude trend's amplitude.  The dipole amplitude remains anomalous (the dashed lines indicate the expected amplitude and direction for the kinematic dipole $\mathcal{D}_{\rm kin}$.) even after inclusion of a general octupole in the inference.}
    \label{fig:declo_decl}
\end{figure}

Beginning with the simplest extension, Fig.~\ref{fig:declo_decl} shows the posteriors of parameters of model (v), marginalised over all octupole angles for clarity, and model (ii).  We infer the following amplitudes:
\begin{align}
    \hat{\mathcal{D}}_{\rm (v)}^{\rm GenPo}=0.0163_{-0.0024}^{+0.0027}\nonumber\\
    \hat{\mathcal{O}}_{\rm (v)}^{\rm GenPo}=0.0247^{+0.0107}_{-0.0110}\\
    \hat{\mathcal{Y}}_{\rm ecl,(v)}^{\rm GenPo}=0.0829_{-0.0036}^{+0.0036}\nonumber
\end{align}
The octupole amplitude $\mathcal{O}$ is inferred to be formally non-zero at $\sim\!2.3\sigma$.  This is consistent with the considerations made in Section~\ref{sec:lowmultipoles_shotnoise}, where we found that an octupole could be detectable at 3 times the noise level ($\sim3\sigma$) only with amplitudes $\mathcal{O}=0.029$ and above; accordingly the octupole seen in Fig.~\ref{fig:declo_decl} does not amount to a significant detection.  Already here we reiterate that finding a numerical value $\mathcal{O}>\mathcal{D}$ does not equate to $C_3>C_1$, and neither does it necessarily amount to a greater modulation of the signal on the sky, cf.~Eq.~(\ref{eq:fO}).\footnote{If one set $\mathcal{O}=\mathcal{A}$ one would find the octupole to result in a minimum-to-maximum variation on the sky of only $[-2/5\cdot\mathcal{A},2/5\cdot\mathcal{A}]$.}  Nevertheless, including the octupole does not decrease the dipole amplitude, instead it increases it from the value found in~(\ref{eq:genpo_decl_dipoleamplitude}) by almost 10\% -- we also observe this correlation in Fig.~\ref{fig:leastsquares2} using least-squares fits -- without inducing much change in the dipole direction. 
The same is observed even if a general quadrupole is considered instead of the linear ecliptic trend, model (vi):
\begin{align}
    \hat{\mathcal{D}}_{\rm (vi)}^{\rm GenPo}=0.0163_{-0.0024}^{+0.0027}\nonumber\\
    \hat{\mathcal{Q}}_{\rm (vi)}^{\rm GenPo}=0.0787_{-0.0040}^{+0.0042}\\
    \hat{\mathcal{O}}_{\rm (vi)}^{\rm GenPo}=0.0245_{-0.0108}^{+0.0108}\nonumber
\end{align}
The bottom section of Table~\ref{tab:evidenceratios} shows the results from these models' Bayesian evidences, where we find no evidence for the inclusion of an octupole given the data of \citet{Secrest:2022uvx}. In fact we find ``substantial'' to ``strong'' evidence''~\citep{Kass:1995loi} in favour of those models that have just a dipole and a modulation capturing the ecliptic latitude trend, over any that also include an octupole.  Lastly, we find no correlation between either quadrupole and octupole or $\mathcal{Y}_{\rm ecl}$ and octupole. Hence it is not surprising that the inclusion of an octupole into the more flexible model~(vii), ${\sf DY}_{\sf ecl}{\sf QO}$, is equally disfavoured in comparison with the corresponding model~(iv), ${\sf DY}_{\sf ecl}{\sf Q}$.

To conclude, the evidence stands against the inclusion of an octupole as seen in Table~\ref{tab:evidenceratios}. With its peak amplitude just below what we defined as the detection level in Section~\ref{sec:lowmultipoles_shotnoise}, a catalogue with more sources (in order to  reduce the shot noise) is required to decide on the orientation and amplitude of the octupole in the number counts of high-redshift quasars. Could it be that the anomalously large dipole amplitude found by \citetalias{Secrest:2020has} and \citetalias{Secrest:2022uvx} was overestimated because of having ignored an octupole?  No -- if anything, it turns out, this would have led to an underestimation of the dipole.

\section{Clustering properties}
\label{sec:clustering}

The description of clustering on various scales is a key step in evaluating the validity of a source catalogue for cosmological studies. This usually equates to testing for agreement between the clustering properties on small scales and those expected in the best-fit cosmology that has been found to describe other data sets well. In the context of the cosmic matter dipole this was first done for the NVSS radio AGN~\citep{Blake:2001bg,Blake:2004dg} which showed good agreement with the expectations of structure formation on small scales. This enabled the study of clustering for resolved/extended sources and raised confidence in the dipole measurement~\citep{Blake:2002gx}. Quasar samples other than the CatWISE2020 catalogue considered here such as the \textit{Quaia} sample~\citep{Storey-Fisher:2023gca}
have also been analysed; the associated photometric redshifts supported analysis of auto- and cross-power spectra on small scales in conjunction with CMB lensing~\citep{Alonso:2023guh,Piccirilli:2024xgo}, as well as analysis of features on large scales~\citep{Alonso:2024emk,Fabbian:2025fdk}.\footnote{However the difficulty of modelling the \textit{Quaia} selection function prevented an accurate measurement of its dipole~\citep{Mittal:2023xub}.}

The CatWISE2020 quasar sample provides the most precise  measurement of the cosmic matter dipole. Following our initial reports (\citetalias{Secrest:2020has} and \citetalias{Secrest:2022uvx}) of the matter dipole anomaly, \citet{Tiwari:2022hnf} presented the power spectrum of the CatWISE2020 catalogue of~\citetalias{Secrest:2020has}, finding good agreement with the $\Lambda$CDM expectation on small scales $\ell\gtrsim20$, after suitably parametrising the redshift distribution and galaxy bias. Here we extend their study by focussing on large scales as well as on clustering and noise properties of the sample that have not been addressed previously. To this end, and also to present an analysis of the larger CatWISE2020 quasar sample of~\citetalias{Secrest:2022uvx}, this section covers the estimation of the angular power spectrum (Section~\ref{sec:clustering_powerspectrum}) which we show is in good agreement with $\Lambda$CDM on all scales \emph{except} the dipole, after the ecliptic latitude trend is properly accounted for (for a discussion of the ecliptic latitude trend's influence on the power spectrum see Appendix~\ref{app:clustering_ecliptic}).


\subsection{Power spectrum estimation and uncertainties}
\label{sec:clustering_powerspectrum}

\begin{figure}
    \centering
    \includegraphics[width=\columnwidth]{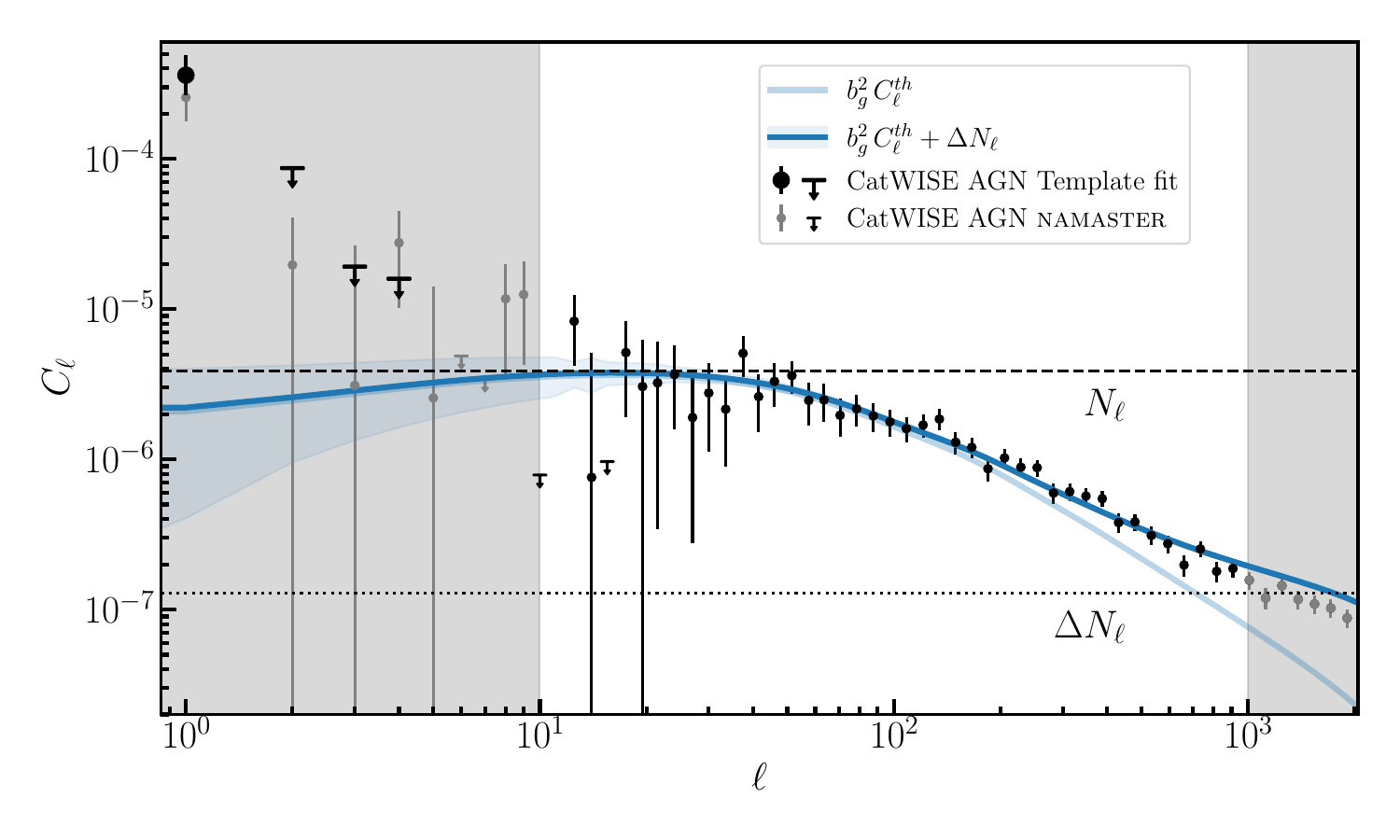}
    \caption{The power spectrum of CatWISE2020 quasars from~\citet{Secrest:2022uvx} after correcting for the ecliptic latitude trend.  Small markers denote the power computed using \textsc{namaster} (with $1\sigma$ errors derived from jackknife sampling after shot noise subtraction), while the large markers are from a template fit for multipoles $\ell \leq 4$ (model (viii), Table~\ref{tab:models} using a Generalised Poisson likelihood, Eq.~(\ref{eq:lnl_genpoisson})).  The solid blue line is a $\Lambda$CDM matter power spectrum fitted to data in the range $10<\ell<1000$ and the blue shaded band indicates cosmic variance.  (The small gray markers are not reliable measurements, therefore not included in the fit, and shown simply for comparison with previous work.)  Dashed and dotted red lines denote, respectively, indicate the level of shot noise and additional ``noise'' due to errors in quasar source finding (e.g.~due to substructure, as discussed in Section~\ref{sec:lowmultipoles_onepoint}).}
    \label{fig:catwise_powerspectrum}
\end{figure}

We first compute the power spectrum of our sample using the pseudo-$C_\ell$ method \citep[e.g.][]{Hivon:2001jp} implemented in \textsc{namaster}~\citep{Alonso:2018jzx}. To do so we create a high resolution $N_{\rm side}=1024$ \textsc{healpix}\footnote{\url{https://healpix.sourceforge.io}}~\citep{Gorski:2004by} map of the source density\footnote{Alternatively one can compute the power spectrum from the source positions themselves without  pixelizing the sky, which can further improve the estimation of the shot noise~\citep{BaleatoLizancos:2023jbr,Wolz:2024dro,Euclid:2024xqh}.} to study the small-scale power. To match this resolution, we redefine the mask of~\citetalias{Secrest:2022uvx} by increasing the masks around point sources to $1^\circ$ radius.\footnote{It is not necessary here to specify a mask apodization scheme; our results are unchanged for a range of such apodizations.}  Subsequently, we remove the source density trend with ecliptic latitude (Eq.~(\ref{eq:selection_ecl})), using the maximum-posterior slope inferred below.\footnote{Using model (ii) in Table~\ref{tab:models}, and a Generalised Poisson likelihood which returned the highest evidence (Table~\ref{tab:evidenceratios}).} We define bins in multipoles with linear spacing of $\Delta\ell=1$ for $\ell<10$, and $\Delta\ell$ increasing logarithmically thereafter until $\ell_{\rm max}=3N_{\rm side}-1$, in $(3N_{\rm side}-1-10)^{0.5}$ steps. The shot noise, determined by the number of sources in the catalogue, here $N=1,595,502$, so $N_\ell=4\pi f_{sky}/N=3.9\times10^{-6}$, has been subtracted from all data points shown in Fig.~\ref{fig:catwise_powerspectrum}.

To estimate uncertainties we repeat the power spectrum estimation on $N_{\rm jk}=400$ jackknife samples (see Appendix~\ref{app:jackknife}) from which we construct the measurement covariance matrix. The error bars shown on each data point $\hat C_\ell$ are taken from the diagonal elements of this matrix. Comparison with the approximate Gaussian covariance~\citep{Garcia-Garcia:2019bku} shows that the jackknife variance is larger for small scales ($\ell>10$), the unshaded area in Fig.~\ref{fig:catwise_powerspectrum}. Hence we retain the larger, jackknife estimate along the diagonal, but replace the off-diagonal elements by the Gaussian approximation for the following step. 

We generate a theoretical ($\Lambda$CDM) matter power spectrum, $C_\ell^{\rm th}$, with \textsc{ccl}~\citep{LSSTDarkEnergyScience:2018yem}, which scales~\citep{Challinor:2011bk} a non-linear $P(k)$ from halofit~\citep{Mead:2020vgs} using the \textit{Planck}-2018 best-fit cosmological parameters~\citep{Planck:2018vyg}, and appropriately accounting for band power coupling and pixel window function. We specify a constant galaxy bias $b_{\rm g}(z,k)=b_{\rm g}=\mathrm{const}$. For the redshift distribution we use a splined $n(z)$ (see Fig.~\ref{fig:redshiftdistribution}) having cross-matched sources in the SDSS Stripe 82 with eBOSS as in~\citetalias{Secrest:2020has}. Our goal here is neither to infer cosmological parameters nor possible forms of $b(z)$ and $n(z)$, but rather to check whether the small-scale power spectrum is consistent with $\Lambda$CDM.  Note that quasars are thought to be more biased with increasing redshift~\citep[cf.~the quadratic fit by][]{Tiwari:2022hnf}.  Considering no redshift dependence of $b_g$ therefore amounts to the possibility of missing some power in small-scale clustering, and overestimating the large-scale clustering power spectrum.  Hence, both render the following results rather conservative.  Lastly, it can be seen from Fig.~\ref{fig:catwise_powerspectrum} that the slope of the fit becomes shallower towards higher $\ell$, an effect that is conventionally modelled by an additional, effective noise contribution, $\Delta N_\ell=\text{const.}$, which we discuss in more detail below.  We  infer the best fitting theoretical power spectrum $C_\ell^{\rm th}$ with two free parameters $b_{\rm g}$ and $\Delta N_\ell$,
\begin{align}
    \hat C_\ell - N_\ell = b_{\rm g}^2C_\ell^{\rm th}+\Delta N_\ell,
    \label{eq:powerfit}
\end{align}
via a Gaussian likelihood using the covariance matrix described above, and considering only the data points between $10<\ell<1000$. This is shown in Fig.~\ref{fig:catwise_powerspectrum}.

We observe that the CatWISE2020 quasar catalogue of \citetalias{Secrest:2022uvx} agrees well with the theoretical power spectrum in the fitted multipole range, even without allowing the bias to be redshift-dependent, which further strengthens the finding by \citet{Tiwari:2022hnf} who used the sample of \citetalias{Secrest:2020has}.  We note that the departure from $b_{\rm g}^2C_\ell^{\rm th}$ (light blue line) at large $\ell \gtrsim 200$ is effectively modelled by the addition of $\Delta N_\ell$ which is due to a number of effects that influence the source count statistics in the sample. It is \emph{not} due to the onset of non-linearity in gravitational clustering as suggested by \citet{Tiwari:2022hnf}, which should not in any case be evident at such  large comoving spatial scales of ${\cal O}(100)$~Mpc. Instead, it is shown in \citet{Tessore:2025auu} that source counts in pixels drawn from distributions that are overdispersed compared with a Poisson distribution lead to an overall increase in power compared with the usual shot noise $N_\ell$. Recalling Section~\ref{sec:lowmultipoles_onepoint}, we had observed precisely this, an effect among other things associated with erroneously identifying different parts of resolved AGN as individual sources, which in turn is observed as an increase in the 2-point correlation on these (small) angular scales~\citep[e.g.][]{Blake:2004dg}.  Using the effective parametrization Eq.~(\ref{eq:powerfit}) we obtain $\Delta N_\ell=1.29\times10^{-7}$ which is about 3.3\% of $N_\ell$. The effective bias is $b_{\rm g}=2.27$, consistent with expectations for mid-IR quasars.

While large scales $\ell \leq 10$ were not part of our fit, we would now like to check if they too agree with the theoretical curve within cosmic variance, or if there is excess large-scale power in our sample on scales other than $\ell=1$. However, the accuracy of pseudo-$C_\ell$ estimates or of their errors (small markers in Fig.~\ref{fig:catwise_powerspectrum}) is not guaranteed on large scales, especially given the presence of a mask as large as the one in use here and the large apparent power in at least the dipole. While we still show these, as small gray markers, for comparison with previous work~\citep[e.g.][]{Tiwari:2022hnf}, our above results allow us to compute posteriors of power in low multipoles that are not prone to named issues. This enables a fair comparison with the large-scale clustering described by the best-fit $C_\ell^{\rm th}$. To this effect, we use the results of Section~\ref{sec:lowmultipoles} and infer real-space templates of those general multipoles that fit the observed number counts on the sky and translate them back into angular power. We intend to use the posteriors of model (vi) in Table~\ref{tab:models} (see Section~\ref{sec:lowmultipoles_nooctupole}) that infers multipole templates up to and including a general octupole. All multipole amplitudes and angles are translated into their corresponding real spherical harmonic coefficient $a_{\ell m}$ for each sample. These are then averaged per multipole to deterministically obtain a posterior of the corresponding $\hat C_\ell$ (see also Section~\ref{sec:lowmultipoles_translation}).  However, since multipoles separated by $\Delta\ell=2$ are coupled by the ecliptic latitude trend (see~Section~\ref{app:clustering_ecliptic}), we anticipate the quadrupole ($\ell=2$) to be correlated with the hexadecapole ($\ell=4$).  We therefore consider an extension to model (vi) that additionally fits for a general hexadecapole, model (viii).  We plot the resulting $C_\ell$ as the larger black markers in Fig.~\ref{fig:catwise_powerspectrum} where the error bars denote the 16th and 84th percentile, and where a downward arrow indicates the 68th percentile if the posterior is consistent with zero.  We observe that among the recovered posteriors for the power at low $\ell$, the only significant departure from the theoretical curve is at $\ell=1$, the dipole. In particular, as elaborated on in Appendix~\ref{app:lowmultipoles_correlation}, the octupole $C_{\ell=3}$ power is found to be constrained in amplitude, to a level at which its influence on the measurement of the dipole is comparably small.

\section{Discussion}
\label{sec:discussion}

We have presented an analysis of spherical harmonic multipoles in the projected number counts of the CatWISE2020 quasar catalogue of \citetalias{Secrest:2022uvx}. We fit low multipoles in real space using templates, and high multipoles in harmonic space through the angular power spectrum. This is to assess biases or potential systematics in our inference of an anomalous matter dipole in this sample~(\citetalias{Secrest:2020has},\citetalias{Secrest:2022uvx}). Our Bayesian inference uses various combinations of low multipole templates and likelihoods tailored to describe best the data at hand. We find that besides a dipole and a trend of source density with ecliptic latitude (approximated by a quadrupole) there is \emph{no} evidence for power in  higher multipoles, in particular an octupole, and the inferred dipole amplitude remains robust to all model variations considered. 

In summary:
\begin{itemize}
    \item The excess dipole in the CatWISE2020 quasar catalogue is not due to bias from power in higher multipoles  (Section~\ref{sec:lowmultipoles_nooctupole} and Fig.~\ref{fig:declo_decl}).
\end{itemize}
This is to put into context \citet{Abghari:2024eja} who put forward the possibility of a thus-far-unmeasured octupole biasing high the measurement of a dipole.  Their argument consisted of sets of simple mocks with artificially injected octupoles of amplitudes large-enough to create a bias in recovered dipole amplitudes.  We respond to their claims more concretely in Appendix~\ref{app:lowmultipoles_correlation} where we contrast their approach with one that is actually informed by the data their mocks attempt to describe.  The more useful procedure is to fit explicitly for a dipole and octupole in the real data (instead of just investigating simulations). We have done this in Section~\ref{sec:lowmultipoles}, finding that the dipole was in fact not overestimated in previous studies of the CatWISE2020 quasar catalogue. The potential mixing of different multipoles in template fits of signals on the masked sky is of course well-known.  In the context of matter dipole measurements, this was already discussed by~\citet{Gibelyou:2012ri} who too performed combined template fits in their analyses of various galaxy samples.

On small scales, our power spectrum analysis demonstrates agreement with expectations from hierarchical structure formation, for a galaxy bias of $b_g\approx 2.25$ as is appropriate for $z\sim1$ quasars; this fit sets the level of clustering on all scales, including the dipole at $\ell=1$, as would be expected in $\Lambda$CDM.  The extrapolation of the best-fit power spectrum to $\ell=1$ directly determines the ``clustering dipole'' contribution to the overall dipole amplitude.  We find:

\begin{itemize}
    \item The $\Lambda$CDM clustering dipole is highly subdominant with respect to the observed dipole (Section~\ref{sec:clustering} and Fig.~\ref{fig:catwise_powerspectrum})
\end{itemize}

In this context, we 
mention the similarities and differences between our power spectrum analysis and that of \citet{Tiwari:2022hnf} using the same catalogue. We confirm their finding that the CatWISE2020 quasar angular power spectrum agrees with the $\Lambda$CDM  expectation on small scales. While our fit did not allow for variation of parameters describing the redshift distribution $f(z)$ or a redshift-dependent galaxy bias $b_g(z)$ (as this was not our objective), we employed jack-knife error bars which are more reliable than assuming Gaussian covariances (see  Fig.~\ref{fig:njk}).

In Section~\ref{sec:lowmultipoles_onepoint} we introduced an effective noise term $\Delta N_\ell$ \citep[][]{Tessore:2025auu} to capture the non-Poissonian overdispersion in the data, which drives the departure of the angular power from its theoretical expectation.  We emphasise that the computation of angular power on large scales for individual multipoles (as well as jack knife errors of the same multipoles) from masked data (gray points at $\ell<10$ in Fig.~\ref{fig:catwise_powerspectrum}) is not recommended. We observe that this power does not  agree with the more principled real-space template fits that led to the large markers at low $\ell$ in the same figure.  Crucially we can directly  estimate the clustering dipole from the extrapolation of the theoretical angular power spectrum template to $\ell=1$, finding $\hat C^{\rm clus}_1=(2.21\pm1.81)\times10^{-6}$ including cosmic variance -- i.e. 140 times smaller than the observed value at $\hat C_{1,{\rm (ii)}}^{\rm GenPo}=(3.10\pm0.75)\times10^{-4}$ (Fig.~\ref{fig:catwise_powerspectrum}). The impact on the observed dipole amplitude can be roughly estimated from $C^{\rm clus}_1$ by sampling randomly oriented dipoles from this power spectrum amplitude and subtracting each from the observed dipole, before querying its amplitude. We thus  find a $1\sigma$ spread of $4.84\%$ around the value inferred here --- quite insufficient to account for the observed anomaly.

Our formulation of general multipole templates (Section~\ref{sec:lowmultipoles_translation}) is not unique and in principle interchangeable with other approaches such as those simply using spherical harmonic templates~\citep[e.g.][]{Gibelyou:2012ri}, or those generated via multipole vectors~\citep[e.g.][]{Copi:2003kt,Oayda:2024voo}. Nevertheless, expanding analytically the respective multipole templates as in Section~\ref{sec:lowmultipoles_translation} has the advantage of negligible computational cost, e.g., when evaluating a likelihood at their parameter values as part of Monte Carlo sampling. Moreover, the expressions for higher multipoles allow us to retain the same formalism usually used for the dipole alone. To enable discussions involving both multipole amplitudes and angular power spectrum amplitudes we provided the necessary relations in Section~\ref{sec:lowmultipoles_translation} which in turn enabled us to define ``shot noise multipoles'' in Section~\ref{sec:lowmultipoles_shotnoise}, thus generalising the concept of a ``shot noise dipole''.

We further studied the noise properties of the sample in terms of its (expected) departure from Poisson statistics ( Section~\ref{sec:lowmultipoles_onepoint}). In order to perform an unbiased inference we justify employing a Generalised Poisson likelihood instead of pure Poisson likelihoods considered so  far~\cite[e.g.][]{Dam:2022wwh,Wagenveld:2023kvi,Wagenveld:2025ewl}. We find virtually no variation in the dipole amplitude in any of the models considered, apart  from slightly broadened posteriors when the overdispersion of the Generalised Poisson likelihood is allowed.

\begin{figure}
    \centering
    \includegraphics[width=\columnwidth]{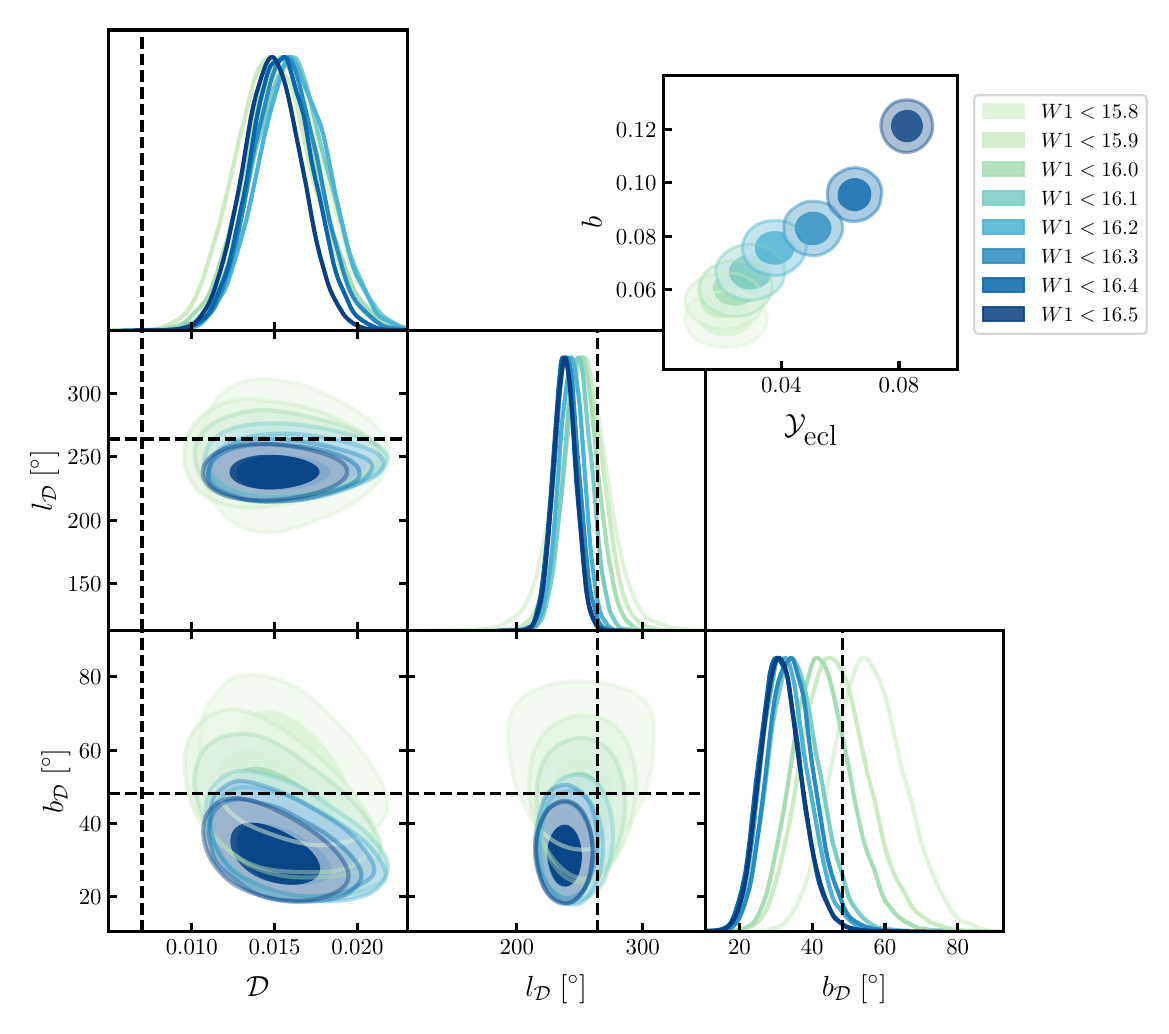}
    \caption{Stability of the dipole amplitude $\mathcal{D}$ across a range of cuts on flux density (our fiducial sample, \citetalias{Secrest:2022uvx}, is listed as ${\rm W1}<16.5$).  Posteriors of the dipole parameters are shown, using model (ii) (${\sf DY}_{\sf ecl}$) and the Generalised Poisson likelihood, Eq.~(\ref{eq:lnl_genpoisson}).  Dashed lines indicate the expected amplitude and direction for the kinematic dipole $\mathcal{D}_{\rm kin}$, which is roughly constant across cuts.  The inset shows the evolution of the amplitude of the ecliptic latitude trend, $\mathcal{Y}_{\rm ecl}$, and the overdispersion parameter, $b$, of the generalised Poisson likelihood, demonstrating that both the ecliptic latitude trend as well as the overdispersion vanish towards brighter magnitudes and hence higher signal-to-noise.}
    \label{fig:fluxcutvariation}
\end{figure}

At this point it is important to note that since our observable remains the source number density, there is another noise property of the data which has not been considered explicitly, viz. significant flux uncertainties. For individual sources this is rare since $>99\%$ of CatWISE2020 sources have observed magnitudes with signal-to-noise higher than 15. However, their presence can give rise to effects such as Eddington bias, which was already suspected by~\citetalias{Secrest:2022uvx} to play a role in the ecliptic latitude trend seen in the data.  While we chose to account for this trend effectively by allowing for it in our inferences via Eq.~\ref{eq:selection_ecl}, we consider it now in more detail to gain a bit more insight into the possible influence of flux uncertainties on our results. Flux uncertainties, and therewith Eddington bias, are expected to decrease with increasing flux cut. A flux-cut-dependent variation of the inferred matter dipole might therefore be suspected if these uncertainties have anything to do with the excess dipole. Fig.~\ref{fig:fluxcutvariation} shows the dipole posteriors, marginalised over all other parameters of model (ii) (cf. Table~\ref{tab:models}), using our fiducial Generalised Likelihood.  With increasing flux cut, all posterior widths increase gradually (due to the concurrently decreasing sample size) and the inferred dipole direction begins to drift slightly towards higher Galactic latitudes $b_{\mathcal{D}}$.  The inset shows separately that restricting the sample to brighter, more certain sources decreases the ecliptic latitude trend's amplitude $\mathcal{Y}_{\rm ecl}$ as would be expected if it was caused by, e.g., Eddington bias.  Curiously, the sample's overdispersion, as described by $b$, is also greatly reduced, which may be useful to identify factors that contribute to the overdispersion of source counts in the first place. The dipole amplitude remains stable across all cuts shown.

\section{Conclusion and Outlook}
\label{sec:outlookconclusion}

There are a number of extensions to our work that would be useful to consider for future analyses. Most importantly, to account for the expected clustering of the CatWISE2020 quasars on all scales, we suggest a pixel-based Bayesian Hierarchical Model (BHM) which simultaneously models the cosmological fluctuations on the sky and their correlations via a power-spectrum-informed pixel covariance, as well as explicit deviations from these on large scales, first and foremost the (potentially kinematic) dipole. This would require constraining a latent Gaussian density field $\delta\hookleftarrow{\cal G}(0,{\bf C}(C_\ell^{\rm th}({\bf p}))$, where parameters ${\bf p}$ determine an expression of the $\Lambda$CDM angular power spectrum $C_\ell^{\rm th}$, which subsequently influences the rate parameter $\lambda_i=\bar{n}_i(1+\delta_i)$ of a (generalized) Poisson-distribution in each pixel $i$ from which observed number counts $d_i$ are drawn as $d_i\hookleftarrow{\rm GenPo}(\lambda_i,b)$.  Low multipole deviations of arbitrary amplitude in addition to the fluctuations modelled here, as well as a selection function, modulate $\bar{n}_i$ as in Eq.~(\ref{eq:numbercounts}). While this was not warranted for the current data (since the higher multipoles considered here $\ell\leq4$ are already subdominant to the dipole, and moreover consistent with noise), this hierarchical approach would be relevant in the analysis of large-scale power in upcoming surveys, where the higher number densities would allow for real space fits at the level expected due to primordial cosmological fluctuations.

In summary, our analysis demonstrates that the anomalously high dipole in the CatWISE2020 quasar catalogue is not an artefact of neglecting to fit higher-multipole power or of mis-modelling the sample's noise properties, nor is it due to a low-redshift `clustering dipole'. Rather it is a robust feature of the data. Even after effectively modelling the sky as a sum of higher multipoles, the dipole amplitude remains roughly twice the expectation based on the standard kinematic interpretation of the CMB dipole. No astrophysical or instrumental effect known to-date can adequately explain this excess.

This finding has profound implications. The standard $\Lambda$CDM model based on the FLRW metric is challenged by this gross mismatch between the matter and CMB dipole. The CatWISE2020 quasar catalogue provides the most significant single-data-set indication of a deviation from cosmic isotropy. If future surveys such as the Legacy Survey of Space \& Time (LSST) by the
Vera C. Rubin Observatory~\citep{LSST:2008ijt}, the \textit{Euclid} satellite~\citep{Euclid:2024yrr}, the Spectro-Photometer for the History of the Universe Epoch of Reionization and Ices Explorer~\citep[SPHEREx;][]{Bock:2025ijl}, and the Square Kilometre Array~\citep[SKA;][]{2019arXiv191212699B} confirm the dipole anomaly, it would signal a breakdown of the Cosmological Principle and invite new physics beyond the $\Lambda$CDM model.

\section*{Acknowledgements}

We thank David Alonso, Dominik Schwarz and Richard Stiskalek for helpful discussions. SvH is supported by a Leverhulme Trust International Professorship Grant to S. Sondhi (No. LIP-2020-014). HD is supported by a Royal Society University Research Fellowship (No. 211046).

This publication makes use of data products from the Wide-field Infrared Survey Explorer, which is a joint project of the University of California, Los Angeles, and the Jet Propulsion Laboratory/California Institute of Technology, funded by the National Aeronautics and Space Administration.

This work made use of the following python packages:
\textsc{harmonic}~\citep{Polanska:2024arc}, \textsc{healpy}~\citep{Zonca:2019vzt}, \textsc{matplotlib}~\citep{Hunter:2007}, \textsc{numpy}~\citep{harris2020array}, \textsc{numpyro}~\citep{Phan:2019elc},  \textsc{scipy}~\citep{Virtanen:2019joe}, and
\textsc{skysegmentor}~\citep{Euclid:2025fby}.

\section*{Data Availability}

The data upon which the majority of this article is based, the CatWISE2020 quasar catalogue of \citetalias{Secrest:2022uvx} is available at \url{https://zenodo.org/records/6784602}.  The full CatWISE2020 catalogue~\citep{2021ApJS..253....8M} is available at, e.g., \href{https://datalab.noirlab.edu/data/catwise}{this URL}.

\bibliographystyle{mnras}
\bibliography{References}

\appendix
\section{Additional figures}
\label{app:additional_figures}

This section collects additional figures referred to in the main text.

\begin{figure}
    \centering
    \includegraphics[width=0.8\columnwidth]{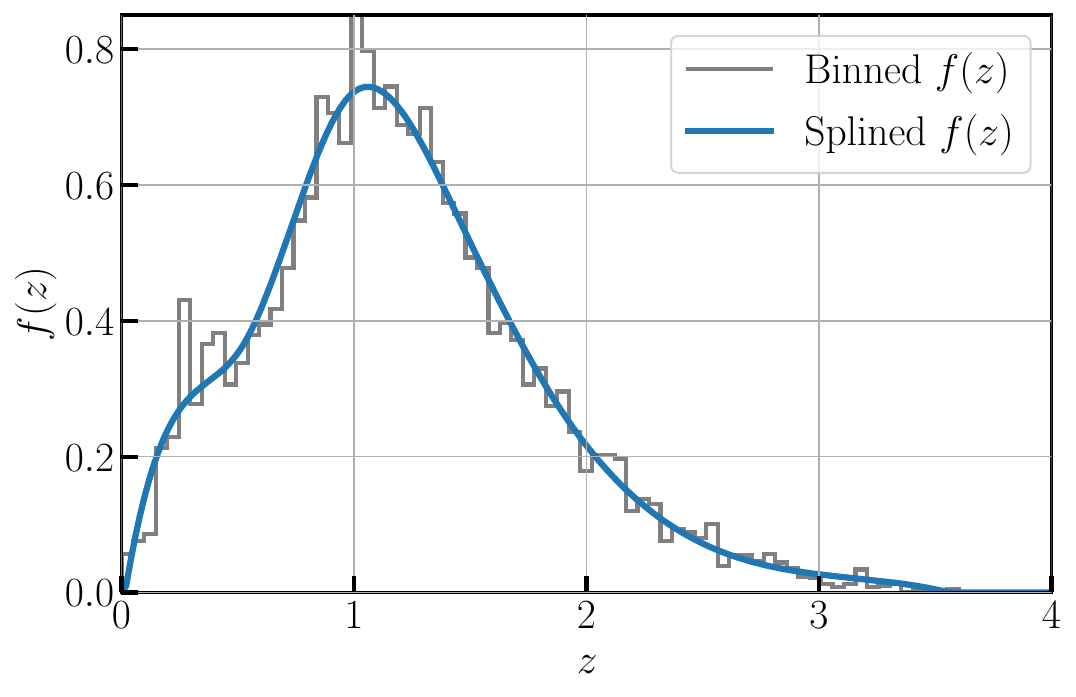}
    \caption{Distribution of spectroscopic redshifts (gray steps) cross-matched between the CatWISE2020 quasar catalogue of \citetalias{Secrest:2022uvx} and eBOSS sources as described in~\citetalias{Secrest:2020has}.  A splined approximation (blue line) to the redshift distribution is used in the computation of the theoretical power spectrum in Section~\ref{sec:clustering_powerspectrum}.}
    \label{fig:redshiftdistribution}
\end{figure}

\begin{figure}
    \centering
    \includegraphics[width=\columnwidth]{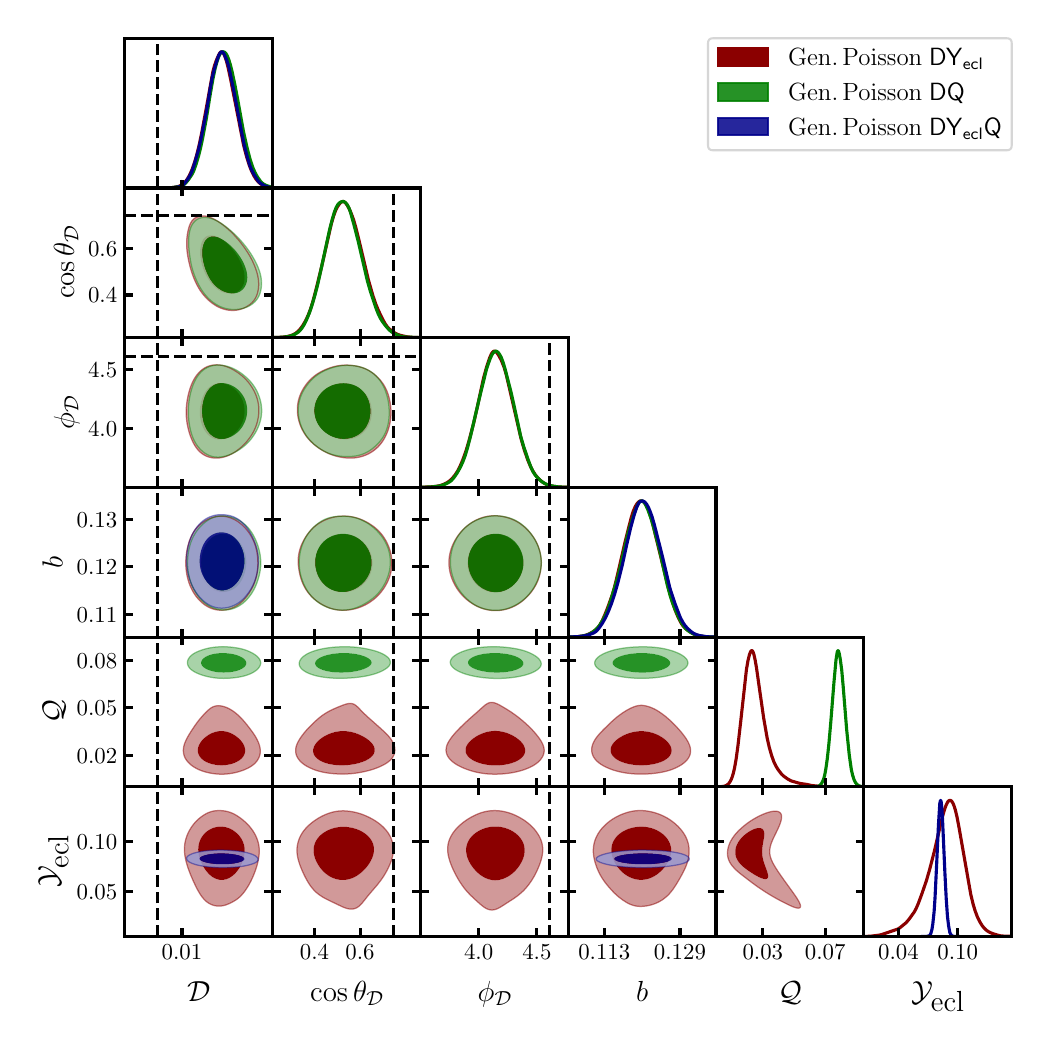}
    \includegraphics[width=\columnwidth]{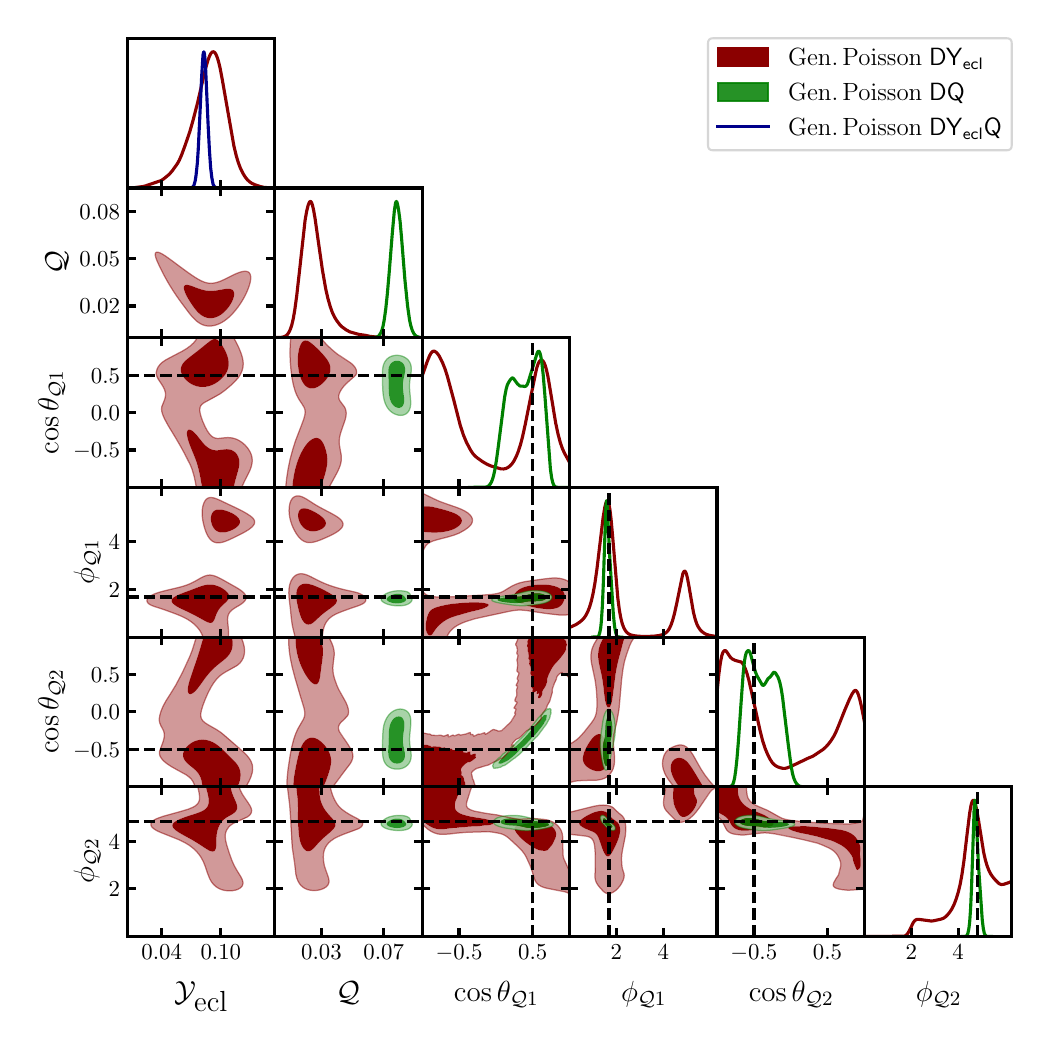}
    \caption{Posteriors of all parameters in models (ii), (iii) and (iv) (${\sf DY}_{\sf ecl}$, ${\sf DQ}$ and ${\sf DY}_{\sf ecl}{\sf Q}$, see Table~\ref{tab:models}) inferred using the Generalised Poisson likelihood (Eq.~(\ref{eq:lnl_genpoisson})), split over two panels for visual clarity.  \textit{Top panel:}  Posteriors were marginalised over all quadrupole angles (except for model (ii)).  The (anomalously large) dipole remains stable across all models (the dashed lines indicate the expected amplitude and direction for the kinematic dipole $\mathcal{D}_{\rm kin}$.) \textit{Bottom panel:}  Posteriors were marginalised over all dipole parameters to highlight the correlation between the ecliptic latitude trend and the general quadrupole.  The quadrupole angles are found to prefer directions close to the ecliptic poles (indicated by dashed lines) even for model (iv).  In this model, the residual quadrupole appears to mainly model deviations from the assumed linearity of the trend with ecliptic latitude.}
    \label{fig:declq_triangle}
\end{figure}

\section{Impact of the ecliptic latitude trend on the power spectrum}\label{app:clustering_ecliptic}

The ecliptic latitude trend found in \citetalias{Secrest:2020has} and \citetalias{Secrest:2022uvx} is well-approximated by a linear trend with absolute ecliptic latitude $|b_{\rm ecl}|$, Eq.~(\ref{eq:selection_ecl}), for the considered magnitude cuts. Because of its symmetry, to first order, this corresponds to a modulation of the  expected number density $\mathcal{N}({\bf n})$ with a quadrupole, while concomitant modulations by hexadecapole, and higher (even) multipoles also contribute, albeit with decreasing amplitudes. This leads to the artificial introduction of these ($\ell=2,4,...$) multipoles on the sky, all aligned with the ecliptic. The number density fluctuations at all scales are modulated by these modes so we investigate how the power spectrum is affected by the ecliptic latitude trend.  In this section we discuss the impact such a modulation has on the angular power spectrum of the unmodulated sky.

For the sake of illustration, consider the modulation of a real signal $f$ on the sphere by an axisymmetric multipole template $t_k=a_{k0}Y_{k0}=\kappa Y_{k0}$, where $Y_{k0}$ is the real spherical harmonic function for multipole $k$ and mode $0$, and $\kappa$ is its amplitude, such that the modulated signal $g=f\, (1+t_k)$.  The estimated power spectrum of $g$ reads
\begin{align}
    \hat C^g_\ell = \hat C^f_\ell +& \frac{2\kappa}{2\ell+1}\sum_m{\rm Re}\left[f^*_{\ell m}\left(f\, t_k\right)^{\phantom *}_{\ell m}\right]\nonumber\\
    +& \frac{\kappa^2}{2\ell+1}\sum_m\left|\left(f\, t_k\right)^{\phantom *}_{\ell m}\right|^2,
\end{align}
where it can be seen that cross-terms between $f$ and $t_k$ contribute to the overall measured power, in addition to the power of the unmodulated signal $\hat C^f_\ell$. The effect of the cross-terms is made more explicit by expressing the spherical harmonic coefficients of the product of $f$ and $t_k$ as
\begin{align}
    (f\, t_k)^{\phantom *}_{\ell m} =&\, \kappa\sum_{\ell'm'}f^{\phantom *}_{\ell'm'}\int{\rm d}\Omega\,Y^{\phantom *}_{\ell'm'}Y^{\phantom *}_{k0}Y_{\ell m}^*\\
    =&\, \kappa\sum_{\ell'}f^{\phantom *}_{\ell'm}\sqrt{\frac{(2\ell+1)(2k+1)(2\ell'+1)}{4\pi}}\nonumber\\
&\qquad\qquad\qquad\times
\begin{pmatrix}
\ell' & k & \ell \\
0 & 0 & 0
\end{pmatrix}
\begin{pmatrix}
\ell' & k & \ell \\
m & 0 & -m
\end{pmatrix}\\
\equiv&\, \kappa\sum_{\ell'}f^{\phantom *}_{\ell'm}{\sf M}_{\ell\ell'm}^k,
\end{align}
where we used the mode sum rule of the Wigner 3$j$ symbols, here $m'-m=0$, in simplifying the sum over $m'$. This allows us to write the estimated power spectrum as
\begin{align}
    \hat C_\ell^g = \hat C_\ell^f+&\frac{2\kappa}{2\ell+1}\sum_m {\rm Re}\left[f^*_{\ell m}\sum_{\ell'}f^{\phantom *}_{\ell'm}{\sf M}^{\phantom *}_{\ell\ell'm}\right]\nonumber\\
    +&\frac{\kappa^2}{2\ell+1}\sum_m \left|\sum_{\ell'}f^{\phantom *}_{\ell'm}{\sf M}^{\phantom *}_{\ell\ell'm}\right|^2.
\end{align}
The second term still arises from the correlation between the modulating template and the underlying function $f$, i.e.~the cross power, while the last term solely describes the mode-coupling, or ``leakage'', of signal from $\ell'$ to $\ell$.  Again due to the properties of the Wigner 3$j$ symbols, the mode coupling matrices ${\sf M}$ are non-zero only for $\ell'=\ell\pm k$ and $\ell'=\ell$. If we consider the ensemble average over different statistically isotropic realizations of $f$, the cross-correlation averages to zero,
\begin{align}
    C_\ell^g = C_\ell^f+\frac{\kappa^2}{2\ell+1}\sum_{\ell'm} C^f_{\ell'}\left|{\sf M}^{\phantom *}_{\ell\ell'm}\right|^2,
\end{align}
which makes the power leakage more apparent.\footnote{Such modulation has a direct relation with the bipolar spherical harmonic formalism, see~\cite{Varshalovich:1988ifq}.}\\

We consider a quadrupolar modulation motivated by the ecliptic latitude trend,\footnote{Comparing with the parameterisation of an axisymmetric quadrupole, Eq.~(\ref{eq:fQ_m=0}), the amplitude $\kappa$ is related to $\mathcal{Q}$ by $\mathcal{Q}=\kappa\sqrt{45/ 16\pi}.$} i.e.~$k=2$, and note that ${\sf M}$ couples power in any two multipoles separated by $\Delta\ell=2$, e.g.~the quadrupole and hexadecapole, or the dipole and octupole. However no coupling is induced between, e.g., dipole and quadrupole. Hence, if the CatWISE2020 quasar catalogue covered the full sky, and if, additionally, the power in octupole and other odd moments is small, the presence of the near-quadrupolar ecliptic latitude trend would have no bearing on the measurement of the dipole. We qualify the assumption behind this statement in Section~\ref{sec:lowmultipoles_nooctupole}, where we indeed find no evidence for any power in the octupole.

The argument above does not necessarily hold when there is also a mask which introduces additional couplings between multipoles of the now modulated map and those of the mask. However, we recall that the mask used for our sample approximately has even symmetry, hence it too will predominantly couple those multipoles that are an even number of $\Delta\ell$ apart. In summary, again using the fact that the octupole power is not significant, the ecliptic latitude trend is not a concern for measurement of the dipole, contrary to the claim by \citet{Abghari:2024eja}.

\begin{figure}
    \centering
    \includegraphics[width=\columnwidth]{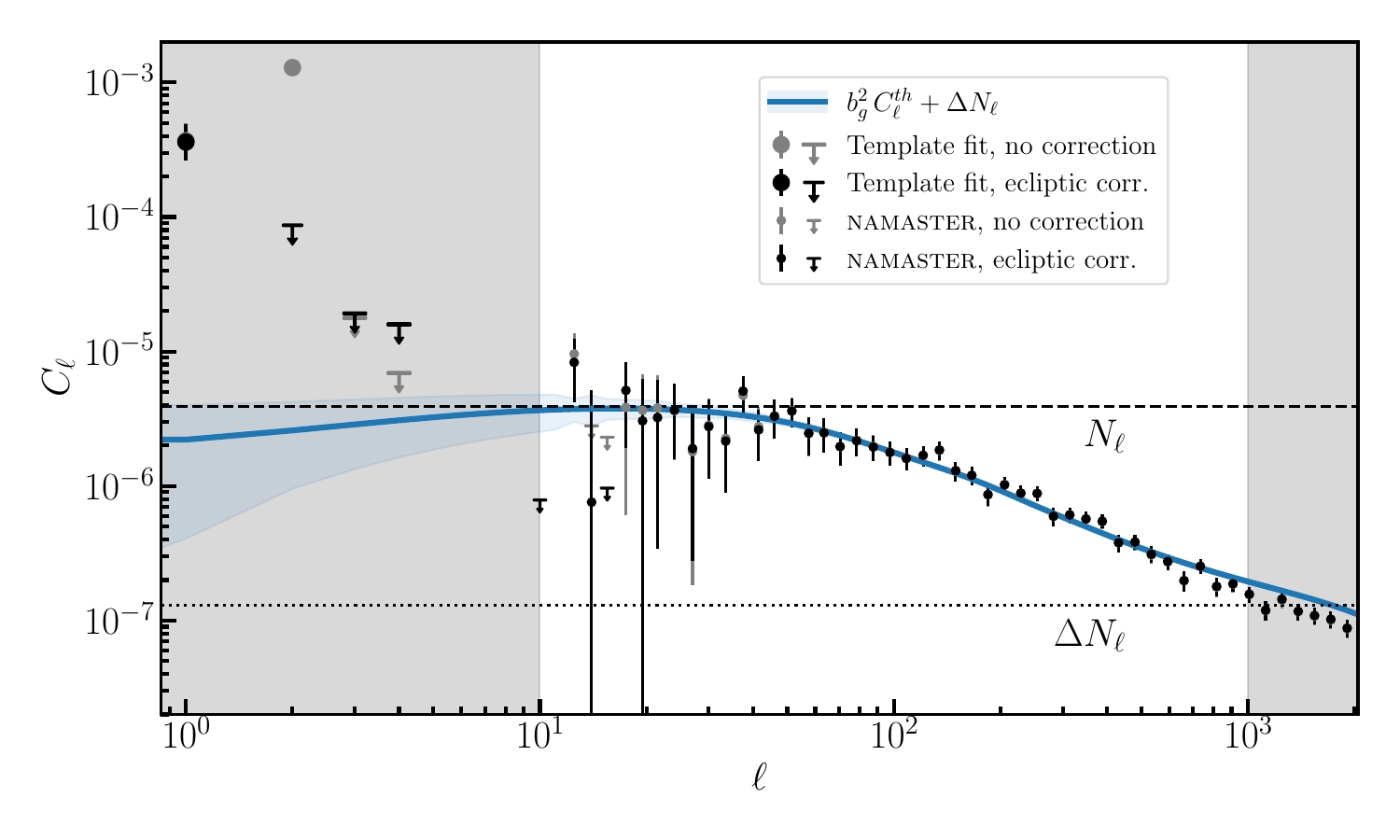}
    \caption{The power spectrum of CatWISE2020 quasars from~\citet{Secrest:2022uvx} before (gray) and after (black) correcting for the ecliptic latitude trend.  For visual clarity jack knife error bars are omitted here.  The power spectrum at small scales ($\ell\gtrsim20$) is unaffected by the ecliptic latitude trend or its correction, reaffirming the observations in Fig.~\ref{fig:catwise_powerspectrum}.  On large scales we employ the inference of multipoles in real space, as in Section~\ref{sec:clustering_powerspectrum}, using model (v), and find the ecliptic latitude trend to affect mostly quadrupole and hexadecapole as expected. The dipole measurement at $\ell=1$ remains robust in all cases.}
\label{fig:catwise_powerspectrum_comparison}
\end{figure}

Fig.~\ref{fig:catwise_powerspectrum_comparison} shows the difference in power spectrum estimates with and without the ecliptic latitude correction, where again we used the maximum-posterior amplitude $\mathcal{Y}_{\rm ecl}$ from Section~\ref{sec:lowmultipoles_nooctupole}, Eq.~\ref{eq:genpo_decl_dipoleamplitude}. Changes are observed mainly at the previously identified large scales ($\ell=2,4$); in particular the quadrupole power drops by an order of magnitude. The dipole ($\ell=1$) is however unchanged.

\section{Jack knife samples}
\label{app:jackknife}

\begin{figure}
    \centering
    \includegraphics[width=\columnwidth]{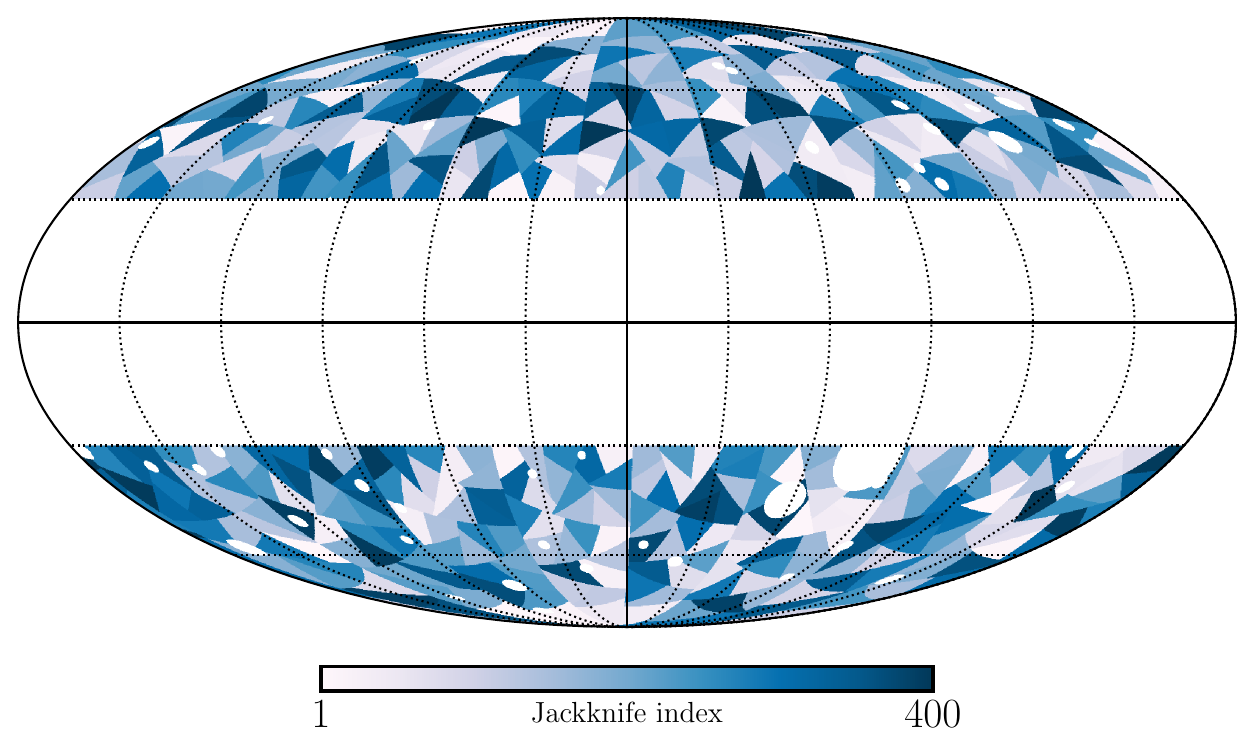}
    \caption{Example of our jackknife tessellations created with \textsc{skysegmentor} and a total number of tiles $N_{\rm jk}=400$.  We undertake the tessellation separately in contiguous patches (one in the Northern and one in the Southern Galactic hemisphere) each with a number of tiles weighted by the relative sky fraction, here $N_{\rm jk}^N=203$ and $N_{\rm jk}^S=197$.}
    \label{fig:jkmap}
\end{figure}

This section provides details on the estimation of uncertainties on the angular power spectrum band powers of Section~\ref{sec:clustering} (Figs.~\ref{fig:catwise_powerspectrum}\&\ref{fig:catwise_powerspectrum_comparison}). Jack knife samples are defined by tessellating the unmasked sky into $N_{\rm jk}$ equal-sized areas\footnote{To better ensure that the segments are indeed close-to-equal in size we use \textsc{skysegmentor}~\citep{Euclid:2025fby} (\url{https://skysegmentor.readthedocs.io/}) but note that estimates using \textsc{kmeans\_radec} (\url{https://github.com/esheldon/kmeans_radec}) return virtually the same results.}, see Fig.~\ref{fig:jkmap}, and excluding one at a time from the power spectrum computation using \textsc{namaster}.

From the jackknife samples we construct the measurement covariance matrix~\citep[e.g.][]{Koukoufilippas:2019ilu},
\begin{align}
    {\rm Cov}(C_\ell,C_{\ell'}) = \frac{N_{\rm jk}-1}{N_{\rm jk}}\sum_i\Delta C_\ell^{(i)}\Delta C_{\ell'}^{(i)},
\end{align}
where $\Delta C_\ell^{(i)}$ is the difference between the power spectrum obtained from the $i$-th jackknife sample's estimate and the average over all $N_{\rm jk}$ power spectra.

\begin{figure}
    \centering
    \includegraphics[width=\columnwidth]{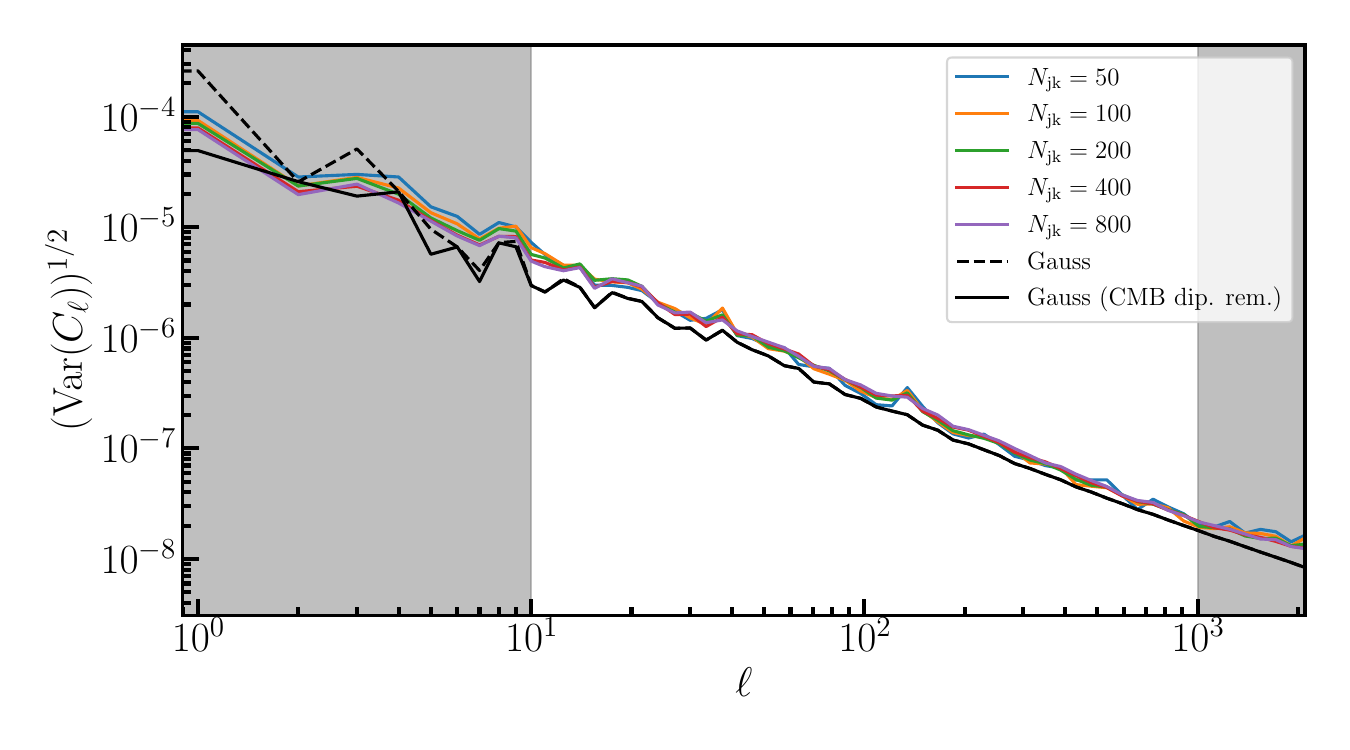}
    \caption{Comparison of jack knife errors depending on the number of segments $N_{\rm jk}$ (colourful lines) and with a Gaussian approximation (black lines).  In the region of interest (unshaded white) we select the more conservative jack knife estimate over the Gaussian approximation with $N_{\rm jk}=400$.}
    \label{fig:njk}
\end{figure}

Fig.~\ref{fig:njk} compares the diagonals of the thus obtained jack knife covariance matrices for different sample numbers $N_{\rm jk}$ (colourful lines).  We find these (as well as the mean deviations from the full sample's power spectrum) to have sufficiently converged for $N_{\rm jk}\gtrsim400$.  The shown errors in Figs.~\ref{fig:catwise_powerspectrum} and~\ref{fig:catwise_powerspectrum_comparison} are therefore computed for $N_{\rm jk}=400$.  We also compare this against a the diagonal of the approximate Gaussian covariance~\citep{Garcia-Garcia:2019bku} (solid black line), finding the latter to exceed the jack knife variance only at low $\ell$, which is further influenced by the presence of the large modulating dipole (see dashed black line that is obtained from having removed the expected kinematic dipole from the data prior to computing the Gaussian covariance, simply for comparison).  However, neither the Gaussian covariance nor the jack knife covariance provide a good estimate of the true errors in the data at these, $\ell\lesssim10$ multipoles, wherefore we resort to computing these via the multipole inference described in the main text.  Lastly, it is known that jack knife covariances can be biased high~\citep{Efron:1981abc} wherefore we attempted to correct this bias via delete-2 samples as suggested in, e.g., \citet{Efron:1981abc, Euclid:2025fby}.  This, however created no discernible difference for $N_{\rm jk}=400$ and we therefore only use the delete-1 samples in our estimate.

\section{Correlation among multipoles on incomplete skies}\label{app:lowmultipoles_correlation}

In Sec~\ref{sec:lowmultipoles_shotnoise} we began to formalise requirements for the evidence of a given multipole's presence in finite number counts of sources, on which we followed-up in Section~\ref{sec:lowmultipoles_nooctupole}, finding no evidence for an octupole in the data.  However, the absence of evidence for a signal notoriously does not equate to the evidence of its absence, and so we must admit the possibility that power from multipoles higher than the dipole is indeed present and might influence the reliability of the dipole's detection via possible correlation (see, e.g., Appendix~\ref{app:clustering_ecliptic}).  Especially the aforementioned sky mask, that mostly serves to cover the Galactic plane, might be thought of as an obstacle in the unbiased recovery of signals on large scales.  Therefore, in this section, we confront this issue in a Frequentist context (to make connection with the analyses originally in \citetalias{Secrest:2020has} and \citetalias{Secrest:2022uvx}) by presenting the results of toy simulations, leading to the understanding that our previous conclusions, i.e.~the establishment of the cosmic dipole anomaly, remain unaffected.  We also motivate the approach of Secs.~\ref{sec:lowmultipoles_inference}\&\ref{sec:lowmultipoles_nooctupole}, in which we present Bayesian inferences of higher multipole signals collectively and formally address to the question about evidence for the absence of higher multipole power in the CatWISE2020 quasar catalogue.

\subsection{Correlations in noise-less simulations}

The question of whether power in other large-scale multipoles can bias the measurement of specifically the dipole reported in \citetalias{Secrest:2020has} and \citetalias{Secrest:2022uvx} was posed recently by \citet{Abghari:2024eja}.  Their aim was to understand the answer to this question by constructing sets of simulations and by studying the impact of sky masks following a choice of different input parameters.  We here revisit their simulations and, also by using the relations presented in the previous sections \ref{sec:lowmultipoles_translation} and \ref{sec:lowmultipoles_shotnoise}, highlight weaknesses in their arguments, e.g., by showing that the input parameters chosen by \citet{Abghari:2024eja} simply fail to realistically describe the data they aim to model.

\begin{figure}
    \centering
    \includegraphics[width=\columnwidth]{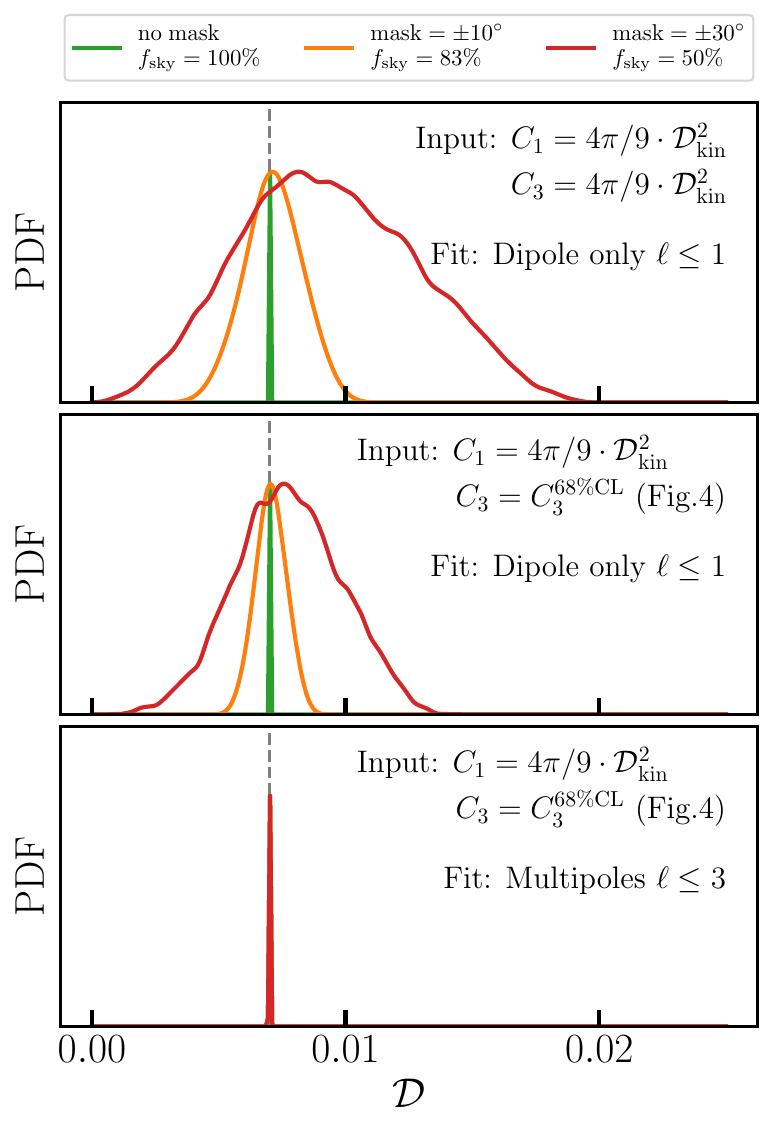}
    \caption{Bias tests using noiseless simulations to study the impact of correlations among multipoles on dipole fits in the presence of masks.  All panels show (smoothed) distributions of recovered dipole amplitudes, $\mathcal{D}$, obtained from random simulations each with $C_1=C_{1,{\rm kin}}$ set by $\mathcal{D}_{\rm kin}=0.007$, Eq.~(\ref{eq:C1}), and different choices of non-zero $C_3$, by using least-squares fits for only $\ell\leq1$ (top panels) and $\ell\leq3$ (bottom panel).\\
    \textit{Top panel:} These simulations follow \citet{Abghari:2024eja} (cf.~their figure 7) incl.~their arbitrary choice to set the octupole power equal to that of the expected kinematic dipole, $C_3=C_{1,{\rm kin}}$. \textit{Middle panel:} Simulations informed by the actually measured octupole power, here represented by the conservative choice $C_3^{68\%{\rm CL}} = 2\times10^{-5}$ (see downward arrow in Fig.~\ref{fig:catwise_powerspectrum}).  \textit{Bottom panel:}  Simulations as in the middle panel, but here the least-squares fit also takes into account the octupole, leading to a near-exact measurement.}
    \label{fig:leastsquares1}
\end{figure}

We first attend to the results presented in their figure 7.  Sets of random sky maps\footnote{We generate sets of $10\,000$ \textsc{healpix} maps with $N_{\rm side}=32$.} are generated, each containing a randomly oriented dipole with amplitude $C_1=C_{1,{\rm kin}}$ matching that of the kinematic expectation $\mathcal{D}_{\rm kin}=0.007$ as per Eq.~(\ref{eq:C1}), and one higher multipole, $\ell>1$, per set, also random in orientation and with an amplitude fixed to the same value $C_\ell=C_{1,{\rm kin}}$.  Subsequently, the skies are cut with Galactic plane masks of different extents, and the maps' dipoles are measured using the same least-squares estimator used in \citetalias{Secrest:2020has}.  \citet{Abghari:2024eja} find a large spread in recovered dipole estimates from their masked simulations if the added multipole $\ell$ is odd, which becomes most pronounced for the octupole, $\ell=3$, as expected (see Appendix~\ref{app:clustering_ecliptic}).  We therefore focus only on the simulations containing octupoles, and, for sake of comparison, reproduce their results in the top panel of Fig.~\ref{fig:leastsquares1}, where it is seen that the randomly  oriented octupoles contribute to apparent dipoles of amplitudes that can exceed the input dipole of $\mathcal{D}_{\rm kin}=0.007$ by more than a factor of 2, given a $\pm 30^\circ$ Galactic plane mask.  While this result is presented as suggestive, it is important to note the choice of octupole amplitude.  Recalling Section~\ref{sec:lowmultipoles_translation}, \citeauthor{Abghari:2024eja}'s choosing the octupole amplitude according to $C_3=C_{1,{\rm kin}}$ appears arbitrary at best, given that the octupole power can actually be constrained with the data the authors claim to describe.  As we found in Section~\ref{sec:lowmultipoles_nooctupole}, the inferred octupole parameters correspond to a posterior in octupole power, $C_3$, that is consistent with zero, and we showed the corresponding 68th percentile as an upper bound in Fig.~\ref{fig:catwise_powerspectrum} (downward arrow at $\ell=3$).  To pick a representative value, we make a rather conservative choice and consider $C_3^{68\%{\rm CL}}=2\times10^{-5}$ as fixed input to the same simulations.  This leads to the results in the middle panel of Fig.~\ref{fig:leastsquares1}, where the spread in recovered dipole estimates is much reduced compared against the previous case, even for the largest mask considered (which is closest to that of the real data).  Given the width of the $C_3$ posterior, it is possible that the sample's true octupole amplitude is different (in fact it is more likely that it is smaller) than that presented here; nevertheless, the conclusions that \citet{Abghari:2024eja} draw based on their figure 7 rely on their exaggerated choice of the simulated octupole amplitude and hence are not generally valid.

We also consider a third variation, where we include higher multipole moments in the least-squares estimator in addition to the dipole fit, as the adverse correlation of an octupole with the dipole would be entirely mitigated by fitting for both multipoles simultaneously.  We present these results for all considered masks and for the case $C_3=C_3^{68\%{\rm CL}}$ in the bottom panel of Fig.~\ref{fig:leastsquares1}, where virtually perfect recovery of the input dipole amplitude is achieved (in fact, the same result would occur if one was to pick $C_3=C_{1, {\rm kin}}$).  This is the avenue we explore with a more dedicated inference on real data in Section~\ref{sec:lowmultipoles_nooctupole}, but first remind the reader that these simulations did not contain any shot noise, wherefore perfect recovery of all simulated multipoles is indeed expected.  As discussed in Section~\ref{sec:lowmultipoles_shotnoise}, shot noise brings with it randomly occurring ``shot noise multipoles'' that may exhibit random correlations with each other, ultimately leading to correlated scatter among the recovered multipole amplitudes.  To address this is the aim of the following simulations.

\subsection{Correlations in simulations including shot noise}

\begin{figure}
    \centering
    \includegraphics[width=\columnwidth]{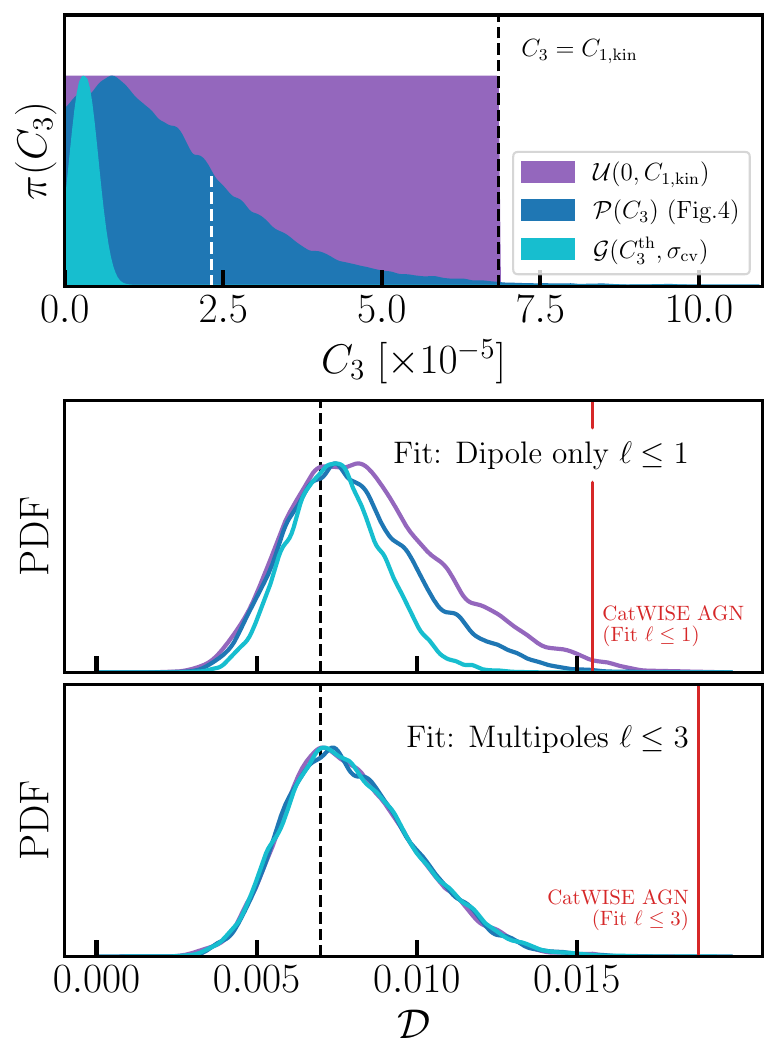}
    \caption{Bias tests using Poisson-sampled simulations to study the impact of correlations among multipoles on dipole fits in the presence of masks.  Dipole amplitudes, $\mathcal{D}$, are recovered from random simulations with $C_1=C_{1,{\rm kin}}$ set by $\mathcal{D}_{\rm kin}$ (Eq.~(\ref{eq:C1})) and $C_\ell$, for $\ell=2,3$, sampled from different prior distributions.\\
    \textit{Top panel:}  Prior distributions on octupole amplitude $C_3$:  The physically unmotivated uniform prior with $C_3\leq C_{1,{\rm kin}}$ (purple, following \citealt{Abghari:2024eja}), the posterior inferred from the actual data as described in Section~\ref{sec:lowmultipoles_nooctupole} (blue; see downward arrow in Fig.~\ref{fig:catwise_powerspectrum} here indicated by dashed white line), and a (truncated) normal prior around the theoretical octupole power $b_{\rm g}^2C_\ell^{\rm th}$ with cosmic variance (cyan; see Fig.~\ref{fig:catwise_powerspectrum}).  \textit{Bottom two panels:}  (Smoothed) distributions of dipole amplitudes, $\mathcal{D}$, recovered from simulations with priors on multipole amplitudes as in top panel (colours matching, see also main text), by using least-squares fits for only $\ell\leq1$ (middle panel) and $\ell\leq3$ (bottom panel).  Red vertical lines indicate the resulting dipole amplitude of the CatWISE2020 quasar catalogue (\citetalias{Secrest:2022uvx}) using the corresponding least-squares fit, in each of the panels, respectively.}
    \label{fig:leastsquares2}
\end{figure}

Shot noise was also introduced by \citet{Abghari:2024eja} in simulations surrounding their figure 8.  There, they present results from simulations of sky maps containing power from combinations of multipoles $\ell\leq3$ on top of a monopole that matches the mean quasar density, which are Poisson sampled in each pixel and subsequently modulated with a dipole of amplitude $\mathcal{D}=\mathcal{D}_{\rm kin}=0.007$ fixed to point in the CMB dipole direction.  The randomly oriented input quadrupoles and octupoles $\ell=2,3$ are sampled from power spectra whose amplitudes are drawn at random from uniform distributions $C_\ell\sim\mathcal{U}(0, C_{1,{\rm kin}})$.  Again dipole amplitudes are measured in these simulations after masking, here with the mask used in our original analysis, and by using the same least-squares estimator as before.  The resulting spread in recovered dipole amplitudes is now due to shot noise (i.e.~``shot noise dipoles'', and correlations with, e.g., ``shot noise octupoles'', etc. as was already part of the simulations in~\citetalias{Secrest:2020has} and~\citetalias{Secrest:2022uvx}), as well as due to the correlations investigated previously in Fig.~\ref{fig:leastsquares1} between the dipole and the additionally input, higher multipoles.  We reproduce their results in the middle panel of Fig.~\ref{fig:leastsquares2} (purple line).  For comparison, we also show the dipole measurement (red vertical line) from the CatWISE2020 quasar catalogue of \citetalias{Secrest:2022uvx}, $\hat{\mathcal{D}}_{\ell\leq1}=0.0155$.  In a similar comparison, \citet{Abghari:2024eja} find that about 140 of their $10\,000$ mocks appear at a dipole amplitude of $0.0155$ or above, leading to a $p$-value of $0.014$; our corresponding simulations reproduce this as well, exhibiting a similar $p$-value of $0.012$.  

As before, \citet{Abghari:2024eja} do not justify their choice of priors for quadrupole and octupole amplitudes that purportedly are ``similar to those that might be in the quasar data set'' \citep{Abghari:2024eja}.  Recalling that mainly the octupole-dipole correlations matter (as opposed to quadrupole-dipole), we illustrate their octupole prior in the top panel of Fig.~\ref{fig:leastsquares2} (purple), where we compare it against two more reasonable prior choices:  a (truncated) normal prior centred on $b_{\rm g}^2C_3^{\rm th}=3\times10^{-6}$ and a width set by cosmic variance, $\sigma_{\rm cv}$ (cyan), and the actually inferred posterior of octupole power (blue) whose 68th percentile was indicated as a downward arrow in Fig.~\ref{fig:catwise_powerspectrum}.  The resulting distributions of recovered dipole amplitudes are shown in corresponding colours in the middle panel of the same figure, formally giving reduced $p$-values of $0.002$ and $<10^{-4}$, respectively.  While one may argue that the prior set to cosmic variance is overly optimistic given the large-scale modes a real data set might contain, we argue that one ought to let the data speak for itself and hence for the inferred (blue) posterior instead of the ad-hoc prior used by \citet{Abghari:2024eja}.  Either way, both $p$-values indicate higher significance than claimed by \citet{Abghari:2024eja}.\\

We close this section with a similar study as previously done in Fig.~\ref{fig:leastsquares1}:  If there truly is a significant octupole in the distributions of sources (whether in a mock sky or the real data) we ought to fit for it explicitly, not to miss the correlation that might exist between it and the dipole we seek to recover.  If the difference between the three distributions observed in the middle panel of Fig.~\ref{fig:leastsquares2} is mainly due to the influence of the different octupoles present in the three sets of simulations, a combined fit for all multipoles $\ell\leq3$ should see this difference vanish.  We demonstrate this explicitly in the bottom panel of Fig.~\ref{fig:leastsquares2}.  The remaining spread in either of the dipole amplitude distributions here is due to the variation of ``shot noise dipoles'' as before, the influence of shot noise provided by higher multipoles than the octupole $\ell>3$, for which we do not explicitly fit, and lastly by a possible correlation between respective pairs of dipoles and octupoles in any one of the simulations.  In fact, such correlation might also be present in the one pair of dipole and octupole present in the real data.  For this reason, and for direct comparison, we also re-compute and show the dipole amplitude found in the sample of \citetalias{Secrest:2022uvx} using the same least-squares fit but now including multipoles $\ell\leq3$ (red vertical line), which places the found dipole amplitude at the larger value of $\hat{\mathcal{D}}_{\ell\leq3}=0.0188$.  While this might be indicative of an even larger dipole amplitude in the distribution of high-redshift mid-IR sources, we save our judgement about the necessity to simultaneously fit an octupole in the data for Section~\ref{sec:lowmultipoles_nooctupole}.  For now it suffices to say, that the $p$-values for all simulations lie at $p<10^{-4}$, formally reinstating the significance of our original detections in \citetalias{Secrest:2020has} and \citetalias{Secrest:2022uvx}.

\bsp
\label{lastpage}
\end{document}